\documentclass[iop,revtex4]{emulateapj}

\shortauthors{MANCUSO ET AL.}
\shorttitle{MAIN SEQUENCES OF GALAXIES AND AGNs}

\begin{document}

\title{The Main Sequences of Starforming Galaxies \\and Active Galactic Nuclei at High Redshift}
\author{C. Mancuso\altaffilmark{1,2,3}, A. Lapi\altaffilmark{1,2,3}, J. Shi\altaffilmark{1,4}, Z.-Y. Cai\altaffilmark{4}, J. Gonzalez-Nuevo\altaffilmark{5}, M. B\'{ethermin}\altaffilmark{6}, L. Danese\altaffilmark{1,2,3}}
\altaffiltext{1}{SISSA, Via Bonomea 265, 34136 Trieste, Italy}
\altaffiltext{2}{INAF-Osservatorio Astronomico di Trieste, via Tiepolo 11, 34131 Trieste, Italy} \altaffiltext{3}{INFN-Sezione di
Trieste, via Valerio 2, 34127 Trieste, Italy}\altaffiltext{4}{Dept. of Astronomy, Univ. of Science and Technology of China, Hefei, 230026 Anhui, China}\altaffiltext{5}{Departamento de F\'isica, Universidad de Oviedo, C. Calvo Sotelo s/n, 33007 Oviedo, Spain}
\altaffiltext{6}{European Southern Observatory, Karl Schwarzschild Stra{\ss}e 2, 85748 Garching, Germany}

\begin{abstract}
We provide a novel, unifying physical interpretation on the origin, the average shape, the scatter, and the cosmic evolution for the main sequences of starforming galaxies and active galactic nuclei at high redshift $z\ga 1$. We achieve this goal in a model-independent way by exploiting: (i) the redshift-dependent SFR functions based on the latest UV/far-IR data from \textsl{HST}/\textsl{Herschel}, and related statistics of strong gravitationally lensed sources; (ii) deterministic evolutionary tracks for the history of star formation and black hole accretion, gauged on a wealth of multiwavelength observations including the observed Eddington ratio distribution. We further validate these ingredients by showing their consistency with the observed galaxy stellar mass functions and AGN bolometric luminosity functions at different redshifts via the continuity equation approach. Our analysis of the main sequence for high-redshift galaxies and AGNs highlights that the present data are consistently interpreted in terms of an \emph{in situ coevolution} scenario for star formation and black hole accretion, envisaging these as local, time coordinated processes.
\end{abstract}


\setcounter{footnote}{0}

\section{Introduction}\label{sec|intro}

Understanding the coevolution of galaxies and supermassive black holes through cosmic times is one of the hottest and most pressing issues of modern research in astrophysics and cosmology.

That some degree of coevolution must be present has been classically established by observing tight relationships between central BH masses and host galaxy properties, such as stellar mass in old stars, luminosity, velocity dispersion, morphological indicators (e.g., Kormendy \& Richstone 1995; Magorrian et al. 1998; Gebhardt et al. 2000; Ferrarese \& Merritt 2000; Tremaine et al. 2002; Marconi \& Hunt 2003; McLure \& Dunlop 2004; Haring \& Rix 2004; Ferrarese \& Ford 2005; Graham 2007; Greene \& Ho 2007; Lauer et al. 2007; Gultekin et al. 2009; Kormendy \& Bender 2009; Vika et al. 2009; Graham et al. 2011; Sani et al. 2012; Beifiori et al. 2012; Kormendy \& Ho 2013; McConnell \& Ma 2013; Ho \& Kim 2014; Shankar et al. 2016), and by recognizing a parallel evolution of the star formation rate (SFR) density for galaxies and of the luminosity density for active galactic nuclei (AGNs; e.g., Boyle \& Terlevich 1998; Franceschini et al. 1999; Heckman et al. 2004; Marconi et al. 2004; Silverman et al. 2009; Madau \& Dickinson 2014; Aird et al. 2015).

From a theoretical viewpoint, $N-$body simulations (e.g., Diemand et al. 2007; Springel et al. 2008; Tinker et al. 2008; Stadel et al. 2009) have been extremely successful in accounting for the large scale matter distribution in the Universe as determined by the primordial dark matter perturbations evolving into bound, virialized structures ('halos') under the action of gravity. However, on smaller, (sub-)galactic scales, the complexity of baryonic physics takes over, making it extremely difficult to provide an ab initio description of all the relevant processes associated to star formation and BH accretion, that occur on vastly different spatial and time scales. This has been demonstrated by the poor predictive capability of current approaches (see Frenk \& White 2012; Scannapieco et al. 2012; see review by Somerville \& Dav\'e 2015) based on hydrodynamic codes (e.g., Vogelsberger et al. 2014; Dubois et al. 2014; Khandai et al. 2015; Schaye et al. 2015; Kaviraj et al. 2016; Richardson et al. 2016) or on (semi-)analytic models (e.g., Bower et al. 2006; Croton et al. 2006, 2016; Fanidakis et al. 2011; Guo et al. 2011; Menci et al. 2012; Somerville et al. 2012, 2015; Lacey et al. 2016).

Such difficulties of theoretical models have originated a longstanding debate concerning the main actors in regulating galaxy and BH coevolution. Three popular scenarios are currently (still) on the market. The first one relies on a prominent role of merging among dark matter halos and associated baryons as the main driver of galaxy and BH evolution; specifically, it envisages merging of gas-rich spirals at high redshift as the main route toward building up massive ellipticals and triggering their star formation and BH activity (e.g., Bower et al. 2006; Croton et al. 2006; Hopkins et al. 2006; Fanidakis et al. 2012; Somerville \& Dav\'e 2015; Guo et al. 2016). An alternative view assumes that star formation and black hole accretion are supported by steady cold gas streams along filaments of the cosmic web (e.g., Dekel et al. 2009; Bornaud et al. 2011). Finally, another view envisages star formation and BH accretion in galaxies to be essentially \emph{in situ}, time-coordinated processes (e.g., Lapi et al. 2006, 2011, 2014; also Lilly et al. 2013; Aversa et al. 2015; Mancuso et al. 2016), triggered by the early collapse of the host dark matter halos, but subsequently controlled by self-regulated baryonic physics and in particular by energy feedback from supernovae (SNe) and AGNs.

The latest interpretation has recently received robust support from
observations of high redshift $z\ga 1$ dusty starforming galaxies, an abundant population discovered via wide areas far-IR/(sub-)mm surveys with \textsl{Herschel}, \textsl{SPT}, \textsl{LABOCA}, and \textsl{SCUBA2}, in many instances thanks to strong gravitational lensing by foreground objects.
Specifically, high-resolution follow-up observations of these galaxies in the far-IR/(sub-)mm/radio band via ground-based interferometers, such as \textsl{SMA}, \textsl{VLA}, \textsl{PdBI} and recently \textsl{ALMA} have revealed star formation to occur in a few collapsing clumps distributed over spatial scales smaller than a few kpc, and at an overall efficiency lower than $20\%$ (e.g., Finkelstein et al. 2013; Neri et al. 2014; Negrello et al. 2014; Riechers et al. 2014; Rawle et al. 2014; Ikarashi et al. 2015; Dye et al. 2015; Ma et al. 2015a; Simpson et al. 2015; Harrison et al. 2016; Scoville et al. 2016; Dunlop et al. 2016).

Moreover, observations of dusty starforming galaxies in the optical and near/mid-IR band from archival data and from the \textsl{Spitzer} space observatory have allowed to characterize their stellar masses. The vast majority of them feature masses strongly correlated to the SFR, in the way of an almost linear relationship dubbed 'Main Sequence', with a normalization steadily increasing as a function of redshift, and with a limited scatter around $0.3$ dex (Daddi et al. 2007; Elbaz et al. 2007; Pannella et al. 2009, 2015; Rodighiero et al. 2011, 2014; Speagle et al. 2014; Whitaker et al. 2014; Renzini \& Peng 2015; Salmon et al. 2015; Tasca et al. 2015; Martis et al. 2016; Erfanianfar et al. 2016; Kurczynski et al. 2016; Tomczak et al. 2016; Schreiber et al. 2016; Dunlop et al. 2016). In addition, the average dust and molecular gas content of main sequence galaxies (Scoville et al. 2014, 2016; B\'ethermin et al. 2015) is found to be consistent with the local, integrated Schmidt-Kennicutt diagram (star formation rate vs. mass of molecular gas). All these findings strongly favor in situ star formation by secular processes over the classical merger-driven scenario, and over streamed gas accretion from cosmological scales. A caveat is that an appreciable fraction of galaxies feature SFR well above the main sequence (Rodighiero et al. 2011, 2015; Silverman et al. 2015), a fact often interpreted as evidence of starbursts triggered by mergers or external inflows; however, we shall show that recent observational evidences on the young age of these systems (e.g., da Cunha et al. 2015; Ma et al. 2015b) point toward an alternative interpretation in line with the in situ scenario.

Recent, model-independent statistical analysis via the continuity equation and the abundance matching techniques (see Peng et al. 2010; Moster et al. 2010, 2013; Behroozi et al. 2013; Aversa et al. 2015; Caplar et al. 2015; Mancuso et al. 2016) have demonstrated that dusty starforming galaxies with SFRs $\ga 10^2\, M_\odot$ yr$^{-1}$ constitute the progenitors of passively-evolving systems with large stellar masses $M_\star\ga 10^{11}\, M_\odot$, that are indeed found to be abundant even at high redshift $z\ga 1$ (see Bernardi et al. 2013; Santini et al. 2012a; Ilbert et al. 2013; Duncan et al. 2014; Tomczak et al. 2014; Caputi et al. 2015; Mawatari et al. 2016; Song et al. 2016). Since massive objects are thought to become passive when their star formation is quenched by the energetic feedback from the central supermassive BH, an exciting bridge between the astrophysics of galaxies and AGNs is naturally established  (see Silk \& Rees 1998; Fabian 1999; King 2003; Granato et al. 2004; Di Matteo et al. 2005; Lapi et al. 2006, 2014; for a recent review see King 2014).

From this point of view, a great impulse in the study of the role played by supermassive BHs in galaxy evolution has come from: (i) X-ray followup observations of AGNs growing at the center of starforming galaxies selected in the far-IR/(sub-)mm or in the K-band (e.g., Borys et al. 2005; Alexander et al. 2005, 2008; Laird et al. 2010; Symeonidis et al. 2010; Xue et al. 2010; Georgantopoulos et al. 2011; Carrera et al. 2011; Melbourne et al. 2011; Rafferty et al. 2011; Mullaney et al. 2012a; Johnson et al. 2013; Wang et al. 2013a; Delvecchio et al. 2015; Rodighiero et al. 2015); (ii) far-IR/(sub-)mm followup observations of the star formation process in galaxies hosting X-ray selected AGNs (e.g., Page et al. 2004, 2012; Stevens et al. 2005; Lutz et al. 2010; Shao et al. 2010; Mainieri et al. 2011; Harrison et al. 2012; Mullaney et al. 2012b, 2015; Rosario et al. 2012; Rovilos et al. 2012; Santini et al. 2012b; Azadi et al. 2015; Barger et al. 2015; Stanley et al. 2015; Harrison et al. 2016) or mid-IR/optically selected quasars (e.g., Carilli et al. 2001; Omont et al. 1996, 2001, 2003; Priddey et al. 2003; Wang et al. 2008a; Walter et al. 2009; Serjeant et al. 2010; Bonfield et al. 2011; Mor et al. 2012; Xu et al. 2015; Netzer et al. 2016; Harris et al. 2016).

These observational studies have revealed a well defined behavior of the average SFR in the host galaxies with respect to the AGN luminosity (see review by Alexander \& Hickox 2012); specifically, the SFR is found to be roughly constant for moderate AGN luminosities, while for high luminosities it stays constant or decreases in X-ray selected AGNs, and increases steeply in mid-IR or optically selected QSOs. A correlation between the average AGN luminosity and the stellar mass emerges also when focusing on mass-selected galaxy samples. All these relationships are often interchangeably referred to as 'AGN main sequence'.

The theoretical interpretation, especially in the range of AGN luminosities investigated via X-ray stacking, is far from trivial;  phenomenological models (Aird et al. 2013; Caplar et al. 2014; Hickox et al. 2014; Stanley et al. 2015) call into play AGN variability, as inspired from the merging scenario (see Di Matteo et al. 2005; Hopkins et al. 2005; Hopkins \& Hernquist 2009; Novak et al. 2011; Hopkins et al. 2013, 2016) and inferred from consistency with the locally observed Eddington ratio distribution. Variability can effectively weaken an underlying correlation between AGN luminosity and SFR, if the AGN luminosity substantially changes (i.e., by more than an order of magnitude) over much shorter timescales than the star formation across the galaxy.

Here we aim at following a different, model-independent approach based on deterministic star formation and BH accretion histories, to provide an unifying physical interpretation on the origin, the shape, the scatter, and the cosmic evolution for the main sequence of both starforming galaxies and AGNs at high redshift $z\ga 1$. To this purpose, we exploit: (i) the redshift-dependent SFR functions, based on the latest UV/far-IR data from \textsl{HST}/\textsl{Herschel}, and related statistics of strong gravitationally lensed sources; (ii) evolutionary tracks for the history of star formation and BH accretion, consistent with a wealth of multiwavelength observations including the observed Eddington ratio distribution at various $z$.  We further validate these ingredients by showing their consistency with the observed stellar mass function of active galaxies, and with the AGN bolometric luminosity functions at different redshifts.

The plan of the paper is as follows: in \S~\ref{sec|STAR} we deal with the main sequence of starforming galaxies, exploiting the SFR functions and deterministic star formation histories to physically interpret its shape, scatter, and cosmic evolution; in \S~\ref{sec|AGN} we follow the same route to interpret the AGN main sequence, exploiting a deterministic BH accretion history consistent with a wealth of multiwavelength observations, and the AGN luminosity functions derived from our SFR functions; in \S~\ref{sec|summary} we summarize and critically discuss our findings.

In the present paper we adopt the flat cosmology indicated by the Planck Collaboration XIII (2016) data, with round parameters: Hubble constant $H_0 = 100\, h$ km s$^{-1}$ Mpc$^{-1}$ for $h = 0.67$, matter density $\Omega_M = 0.32$, baryon density $\Omega_b = 0.05$, and mass variance $\sigma_8 = 0.83$ on a scale of $8\, h^{-1}$ Mpc. Galaxy stellar masses and luminosities (or SFRs) refer to the Chabrier's (2003) initial mass function (IMF).

\section{The main sequence of starforming galaxies}\label{sec|STAR}

Our analysis relies on two basic ingredients: a model-independent determination of the SFR functions at different redshifts; the time dependence of the star formation rate within high-redshift starforming galaxies, as inferred from observations and supported by simple physical arguments. We now describe these two ingredients in some detail, and then investigate the implications for the main sequence of starforming galaxies.

\subsection{Star formation rates functions}\label{sec|SFRfunc}

Our starting point is the global SFR function ${\rm d}N/{\rm d}\log \dot M_\star$, namely the number density of galaxies per logarithmic bin of SFR $[\log \dot M_\star,\log\dot M_\star+{\rm d}\log\dot M_\star]$ at given redshift $z$. This has been accurately determined on the basis of the most recent far-IR and UV data by Mancuso et al. (2016). We defer the interested reader to that paper for a detailed description of the procedure and of the extensive validation against independent datasets; here we recall some basic notions to be used in the sequel.

The SFR function has been built up by exploiting the most recent determinations of the luminosity functions at different redshifts from far-IR and UV data, with the latter being dust-corrected according to the $\beta_{\rm UV}$-IRX relation (see Meurer et al. 1999; Calzetti et al. 2000; Bouwens et al. 2009, 2015, 2016). Specifically, in Mancuso et al. (2016; cf. their Figs.~1 and 2 and Table 1) we adopted a Meurer/Calzetti extinction law, while in the present paper we switch to a Small Magellanic Cloud (SMC) extinction law that better describes the IR excess of UV selected galaxies at $z\ga 2$ (see Bouwens et al. 2016), especially at low SFR $\dot M_\star\sim$ a few tens $M_\odot$ yr$^{-1}$. We note that the determination of the SFR functions is only marginally affected by the difference between the SMC and the Meurer/Calzetti extinction laws. The luminosity $L_{\rm SFR}$ is converted into the SFR $\dot M_\star$ using $\log {\dot M_\star/ M_\odot~{\rm yr}^{-1}} \approx -9.8+\log {L_{\rm SFR}/ L_\odot}$, a good approximation both for far-IR and (intrinsic) UV luminosities under the assumption of a Chabrier's IMF (see Kennicutt \& Evans 2012).

\begin{figure}
\epsscale{1}\plotone{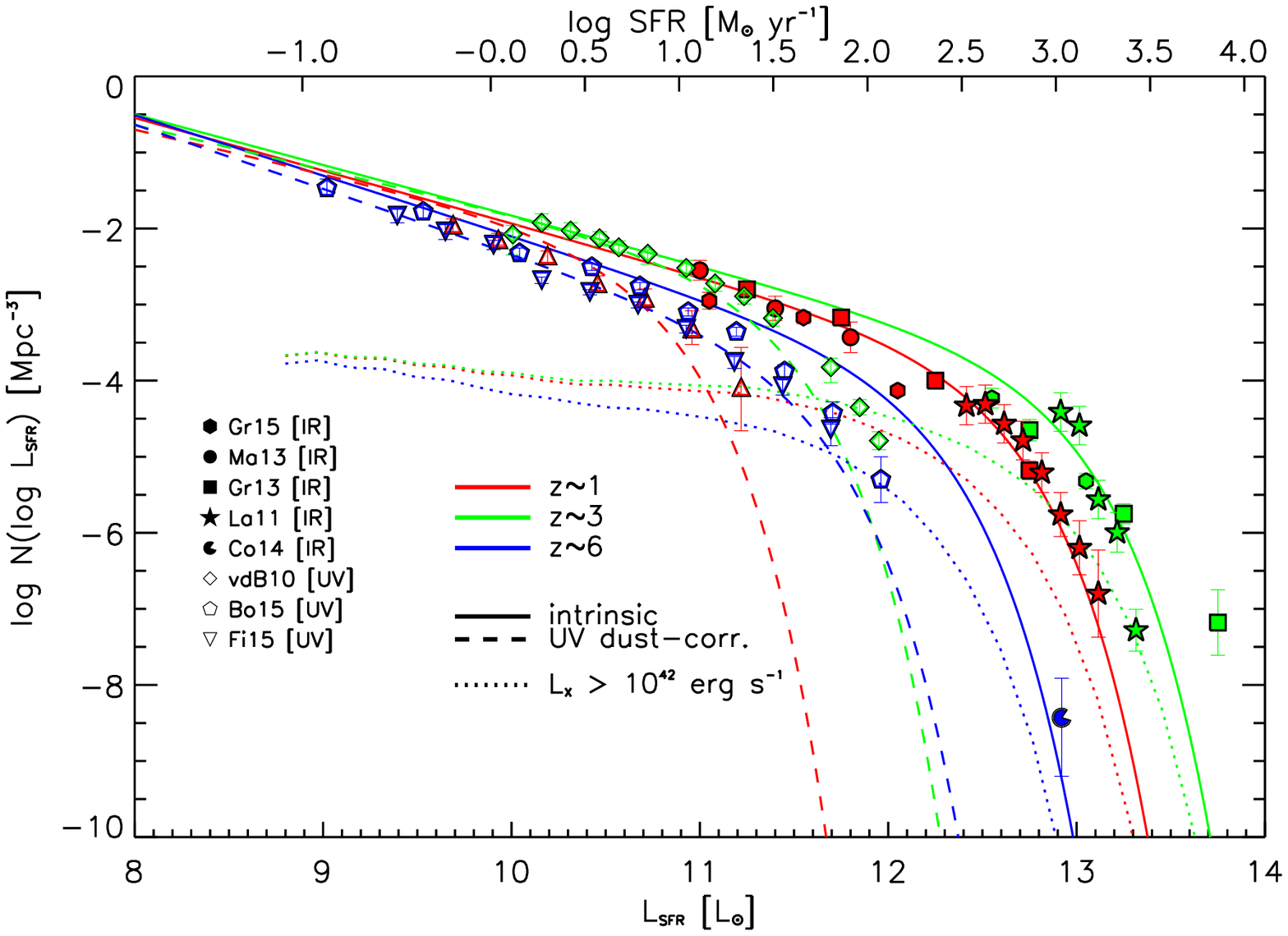}\caption{The SFR functions at redshifts $z=1$ (red lines), $3$ (green lines) and $6$ (blue lines) determined according to the procedure by Mancuso et al. (2016). Solid lines refer to the global SFR function based on UV+far-IR measurements, while dashed lines to the one based solely on UV measurements (dust-corrected via the UV slope, see \S~\ref{sec|SFRfunc} for details). Dotted lines are the SFR functions of galaxies hosting an AGN with X-ray luminosity larger than $10^{42}$ erg s$^{-1}$. UV data (open symbols) are from van der Burg et al. (2010; diamonds), Bouwens et al. (2015; pentagons) and Finkelstein et al. (2015; inverse triangles); far-IR data from Gruppioni et al. (2015; hexagons), Magnelli et al. (2013; circles), Gruppioni et al. (2013; squares), Lapi et al. (2011; stars), and Cooray et al. (2014; pacmans).}\label{fig|SFR_func}
\end{figure}

Then a widely used, smooth analytic representation of the SFR function is found in terms of a standard Schechter shape
\begin{equation}
{{\rm d}N\over {\rm d}\log\dot M_\star}(\dot M_\star,z) = \mathcal{N}(z)\, \left[\dot M_\star\over\dot M_{\star, c}(z)\right]^{1-\alpha(z)}\,e^{-\dot M_\star/\dot M_{\star, c}(z)}~;
\end{equation}
the redshift evolution for any parameter $p(z)$ of the Schechter function, i.e. the normalization $\mathcal{N}$, the characteristic SFR $\dot M_{\star, c}$ and the faint end slope $\alpha$, has been described as a third-order polynomial in log-redshift $p(z)=p_0+p_1\, \xi+p_2\,\xi^2+p_3\,\xi^3$ with $\xi=\log(1+z)$. The values of the evolution parameters $\left\{p_i\right\}$ have been set by performing an educated fit to the data. Specifically, for redshift $z\la 3$ UV data are fitted for SFRs $\dot M_\star\la 30\, M_{\odot}$ yr$^{-1}$ since in this range dust-corrections based on the $\beta_{\rm UV}$-IRX ratio are reliable, while far-IR data are fitted for SFRs $\dot M_\star\ga 10^2\, M_{\odot}$ yr$^{-1}$ since in this range dust emission is largely dominated by molecular clouds and reflects the ongoing SFR. On the other hand, for $z\ga 8$ the (dust-corrected) UV data are considered reliable estimators of the intrinsic SFR function, since the amount of dust in a star-forming galaxy is expected to be rather small for an age of the Universe shorter than $6\times 10^8$ yr. With these anchor points, we interpolate the behavior of the SFR function at intermediate redshifts $z\sim 4-8$, where sampling by far-IR surveys is absent due to sensitivity limits of current instruments. The values of the fitting parameters are reported in Table 1 and the resulting SFR function for three representative redshifts $z\approx 1$, $3$, and $6$ is illustrated in Fig.~\ref{fig|SFR_func}.

All in all, at $z\ga 4$ our estimate implies a significant number density of dusty starforming galaxies with SFR $\dot M_\star\ga 10^2\, M_{\odot}$ yr$^{-1}$, currently missed by UV data. To highlight more clearly this point, we also report in Fig.~\ref{fig|SFR_func} the SFR function that would have been inferred basing solely on UV data, dust corrected via the UV slope. These strongly underestimate the global SFR function for SFRs $\dot M_\star\ga 30\, M_\odot$ yr$^{-1}$. This is because violent SFRs occur within heavily dust-enshrouded molecular clouds, while the UV slope mainly reflects the emission from stars obscured by the diffuse, cirrus dust component (see Silva et al. 1998; Coppin et al. 2015; Reddy et al. 2015).

In Mancuso et al. (2016) we have validated the SFR functions against independent datasets, including galaxy number counts at significative submm/far-IR wavelengths, redshift distributions of gravitationally lensed galaxies, cosmic infrared background.

\begin{figure}
\epsscale{1}\plotone{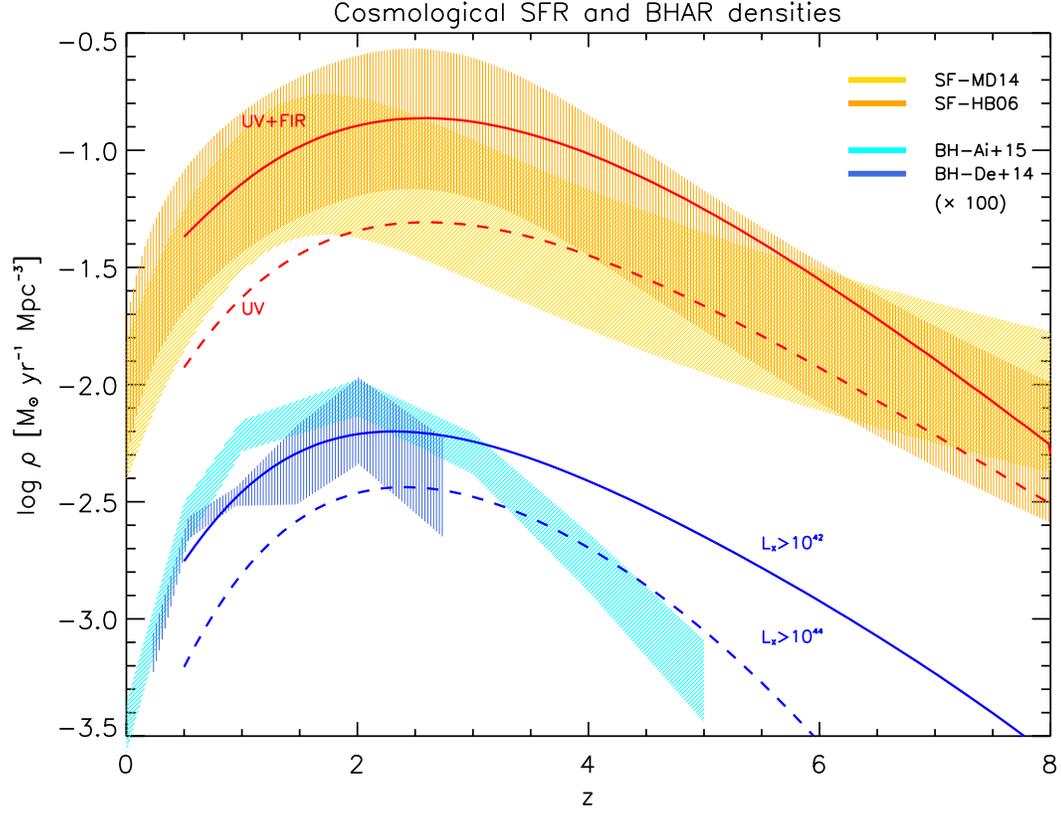}\caption{Cosmic evolution of the average SFR and BH accretion rate density as a function of redshift. Red lines illustrate the cosmic SFR density obtained from integrating our SFR functions, with the solid line referring to the UV+far-IR ones, and the dashed line to the purely UV-inferred ones. Data are from Madau \& Dickinson (2014; yellow hatched area) and Hopkins \& Beacom (2006; orange hatched area). Blue lines illustrate the cosmic BH accretion rate density (multiplied by a factor $100$) obtained from integrating our AGN bolometric luminosity functions (adopting a reference radiative efficiency of $0.1$) down to $L_X\ga 10^{42}$ (solid line) and $10^{44}$ erg s$^{-1}$ (dashed line). Data are from Aird et al. (2015; cyan hatched area) and Delvecchio et al. (2014;  blue hatched area).}\label{fig|cosm}
\end{figure}

In Fig.~\ref{fig|cosm} we also illustrate the cosmic SFR density, computed as
\begin{equation}
\rho_{\dot M_\star}(z) = \int{\rm d}\log \dot M_{\star}\, {{\rm d}N\over {\rm d}\log \dot M_\star}\, \dot M_\star~,
\end{equation}
integrated as usual down $0.03\, \dot M_{\star,c}(z)$ for fair comparison with data. The UV+far-IR result well compares with the observational estimate by Hopkins \& Beacom (2006) based on multiwavelength data including radio. The UV-inferred result is appreciably lower especially at $z\la 6$, and agrees better with the estimate by Madau \& Dickinson (2014) mainly based on UV data dust corrected via the UV slope.

\subsection{Star formation history}\label{sec|STAR_history}

We now focus on the star formation history $\dot M_\star(\tau|M_\star,t)$; this quantity represents the behavior of the SFR $\dot M_\star$ as a function of the internal galactic age $\tau$ (i.e., the time since the beginning of significant star formation) for a galaxy with stellar mass $M_\star$ at cosmological time $t$ (corresponding to redshift $z$).
We base on the indications emerging from many studies of SED-modeling for high $z\ga 2$ starforming galaxies (e.g., Papovich et al. 2011; Smit et al. 2012; Moustakas et al. 2013; Steinhardt et al. 2014; Citro et al. 2016; Cassar\'a et al. 2016); these suggest a slow power-law increase of the SFR $\dot M_\star$ over a timescale $\tau_{\rm b}$, then followed by an exponential decline with timescale $\tau_{\rm SFR}$; in the literature a similar time evolution is sometimes referred to as `delayed exponential model' (see Lee et al. 2009). Such an overall behavior can be described as follows
\begin{eqnarray}\label{eq|STARlightcurve}
\nonumber \dot M_\star(\tau|M_\star,t) &=& \dot M_{\star,{\rm b}}\, (\tau/
\tau_{\rm b})^\kappa ~~~~ 0 \le \tau \le\tau_{\rm b}\\
\\
\nonumber &=& \dot M_{\star, {\rm b}}\, e^{-(\tau-\tau_{\rm b})/\tau_{\rm SFR}}  ~~~~  \tau\ge \tau_{\rm b}\\
\nonumber
\end{eqnarray}
with $\kappa\approx 0.5$; the value of the SFR $\dot M_{\star, {\rm b}}$ at $\tau_b$ is easily related to the final stellar mass $M_\star$ by the relation $\dot M_{\star, {\rm b}}=M_\star\, [\tau_{\rm b}/(\kappa+1)+\tau_{\rm SFR}]^{-1}$.

As to the parameters involved in the above expressions, recent observations by \textsl{ALMA} have shown that in massive high-redshift galaxies the star formation occurred over timescales $\tau_{\rm b}\la 0.5-1$ Gyr at violent rates $\dot M_\star\ga$ a few $10^2\, M_\odot$ yr$^{-1}$ in heavily dust-enshrouded environments (e.g., Scoville et al. 2014, 2016).

A duration of the main star formation episode $\tau_{\rm b}\la 0.5-1$ Gyr in massive high-redshift galaxies, which are thought to be the progenitors of local ellipticals, is indeed confirmed by observations of the $\alpha-$enhancement, i.e., iron underabundance compared to $\alpha$ elements. This occurs because star formation is stopped, presumably by AGN feedback, before type I$a$ SN explosions can pollute the interstellar medium with substantial iron amounts (e.g., Thomas  et al. 2005; Gallazzi et al. 2006; for a review see Renzini 2006). Contrariwise, in low-mass galaxies with $M_\star\la 10^{10}\, M_\odot$ data on the age of stellar population and on chemical abundances indicate that star formation has proceeded for longer times, regulated by type II SN feedback and galactic fountains (see reviews by Conroy 2013 and Courteau et al. 2014).

On this basis, following Aversa et al. (2015) we parameterize the timescale for the duration of the main starformation episode as
\begin{eqnarray}\label{eq|tau_burst}
\nonumber \tau_{\rm b} &=& 1\, {\rm Gyr}\,\left({1+z\over 3.5}\right)^{-3/2}\times \\
\\
\nonumber &\times& \left[1+2\,{\rm erfc}\left({4\over 3}\,\log{\dot M_\star\over 5\,
M_\odot~{\rm yr}^{-1}}\right)\right]~.
\end{eqnarray}
The dependence on cosmic time matches that of the dynamical time $t_c\propto 1/\sqrt{G\,\rho}\propto (1+z)^{-3/2}$, in turn following the increase in average density $\rho\propto (1+z)^3$ of the ambient medium.

As to the quenching timescale $\tau_{\rm SFR}$, the observed fraction of FIR-detected host galaxies in X-ray (e.g., Mullaney et al. 2012b; Page et al. 2012; Rosario et al. 2012) and optically selected AGNs (e.g., Mor et al. 2012; Wang et al. 2013b; Willott et al. 2015) points toward a SFR abruptly stopping, at least in massive galaxies, after $\tau_{\rm b}$ over a short timescale $\tau_{\rm SFR}\la 10^8$ yr due to the action of AGN feedback.

In Fig.~\ref{fig|lightcurves} we show the ensuing star formation and BH accretion histories as a function of the galactic age. We test the adopted star formation history and timescales by connecting the SFR functions to the stellar mass function via the continuity equation. In the absence of merging terms, the continuity equation can be written as
\begin{eqnarray}\label{eq|continuity}
\nonumber {{\rm d}N\over {\rm d}\log \dot M_\star}(\dot M_\star,t) &=& \int{\rm d}\log M_\star~\partial_t \left[{{\rm d}N\over {\rm d}\log M_\star}(M_\star,t)\right]\times \\
\\
\nonumber &\times& {{\rm d}\tau\over {\rm d}\log \dot M_\star}(\dot M_\star|M_\star,t)~;
\end{eqnarray}
here the term on the l.h.s. is the (known) SFR function, while under the integral on the r.h.s. the first term is the cosmic time derivative of the (unknown) stellar mass function, and the second is the time spent by a galaxy in a bin of SFR obtained from the star formation history after Eq.~(\ref{eq|STARlightcurve}). We solve the continuity equation to derive the stellar mass function along the lines discussed by Aversa et al. (2015) and Mancuso et al. (2016). In Fig.~\ref{fig|STAR_MF} we compare the resulting stellar mass function to the recent observational data at various redshifts, finding an excellent agreement.

This not only further substantiates our star formation history and timescales, but also lends strong support toward an in situ coevolution scenario of galaxy formation (e.g., Granato et al. 2004; Lapi et al. 2006, 2011, 2014; Lilly et al. 2013; Aversa et al. 2015). This envisages star formation in galaxies to be mainly a local process regulated by energy feedback from SNe and from the central supermassive BH. In the early stages  the SFR is regulated by SN feedback and slightly increases with time, while the AGN luminosity rises exponentially. After a fraction of Gyr in massive galaxies the nuclear power becomes dominant, removing gas and dust from the interstellar medium and quenching abruptly star formation. Thereafter the stellar populations evolve passively and the galaxy becomes a 'red and dead' early-type.

We note that adopting a conceivable scatter $0.15$ dex around the average star formation timescale changes only marginally the mass function at the high mass end. On the other hand, basing on the (dust-corrected) UV-inferred SFR function leads to strongly underpredict the high mass end of the stellar mass function; this just reflects the undersampling of galaxies with high SFRs by UV data (cf. Fig.~\ref{fig|SFR_func}).

\subsection{Interpreting the galaxy main sequence}\label{sec|STAR_MS}

Given the ingredients above, we populate the SFR vs. $M_\star$ diagram as follows. The number of galaxies per logarithmic bins of SFR and stellar mass is given by
\begin{equation}\label{eq|MS}
{{\rm d^2} N\over {\rm d}\log\dot M_\star\,{\rm d}\log M_\star} \simeq {{\rm d} N\over {\rm d}\log\dot M_\star}\,{{\rm d}\delta\over {\rm d}\log M_\star}~;
\end{equation}
this expression is actually an excellent approximation holding when the star formation $\dot M_\star(\tau)\propto \tau^\kappa$ increases slowly with the galaxy lifetime, and specifically milder than $\kappa\la 1$; we recall that here $\kappa\approx 0.5$ is fiducially adopted on the basis of observations (cf. \S~\ref{sec|STAR_history}). In the above expression the factors on the r.h.s. are the SFR functions (cf. \S~\ref{sec|SFRfunc}) and the relative time $\delta$ spent by the starforming galaxy in a given logarithmic bin of $M_\star$; according to the star formation history (Eq.~\ref{eq|STARlightcurve}) the latter just reads
\begin{equation}\label{eq|Mstar_duty}
{{\rm d}\delta\over {\rm d}\log M_\star}={M_\star\over \dot M_\star}\, {1\over \tau_{\rm b}+\xi\, \tau_{\rm SFR}}\,\ln 10~.
\end{equation}
Here the total duration of the star formation period is taken as $\tau_{\rm b}+\xi\, \tau_{\rm SFR}$ with $\xi\approx 3$, since after that time the stellar luminosity is already decreased by a factor $e^{-\xi}\la 0.05$
and the stellar mass has already attained its final value to a very good approximation; anyway, variations of this parameter do not affect appreciably our results.

The SFR vs. $M_\star$ diagram at the representative redshift $z\approx 2$ is presented in Fig.~\ref{fig|MS_z2}, where the color-code indicates the logarithmic number density of galaxies per unit comoving volume (in Mpc$^{-3}$) after Eq.~(\ref{eq|MS}). The lilac line with errorbars illustrates the number density-weighted mean relationship $\langle\dot M_\star\rangle$ at given $M_\star$ with its $2\sigma$ variance; this is the so called 'main sequence' of starforming galaxies. We stress that averaging over the SFR function weighted by the relative time spent at given $M_\star$ is equivalent to perform a mass selection. In this respect, our outcome well compares with the observational determinations based on statistics of large multiwavelength (UV+IR), mass-selected samples (white shaded areas) by Rodighiero et al. (2011, 2014) and by Speagle et al. (2014).

We remark that the main sequence originates naturally in our approach as a statistical locus in the SFR vs. $M_\star$ plane. However, this by no means implies that an individual galaxy climbs along the main sequence during its lifetime. Typical evolutionary tracks followed by individual objects, as inspired by the in situ coevolution scenario (cf. \S~\ref{sec|STAR_history}), are illustrated by dotted lines; their shape is dictated by the slowly increasing SFR $\dot M_\star\propto \tau^{1/2}$ and appreciably rising stellar mass $M_\star\propto \tau^{3/2}$, which imply $\dot M_\star\propto M_\star^{1/3}$. Then the main sequence with its associated variance correspond to the portions of such tracks where galaxies spend most of their lifetime in logarithmic bins of $M_\star$, see Eq.~(\ref{eq|Mstar_duty}).

To highlight the relevance of observational selections different from the one based on stellar mass, in Fig.~\ref{fig|MS_z2} we also report data points for individual, far-IR selected galaxies by Koprowski et al. (2016), Ma et al. (2015b), Negrello et al. (2014) plus Dye et al. (2015), and da Cunha et al. (2015) mainly at redshifts $1\la z\la 3$. We note that the observations by Koprowski et al. are drawn from a \textsl{SCUBA2} survey area of $10^2$ arcmin$^2$, while the other data are extracted from \textsl{Herschel} survey areas of $10^2$ deg$^2$; as a consequence, the former sample can probe galaxies with SFRs of few $10^2\, M_\odot$ yr$^{-1}$ at most, while the latter samples can probe galaxies with more extreme SFR values up to few $10^3\, M_\odot$ yr$^{-1}$, whose number density is substantially lower. On the other hand, all these far-IR samples are limited by instrumental sensitivity to a minimum SFR around $10^2\, M_\odot$ yr$^{-1}$. We also report data from the recent ALMA observations of the \textsl{Hubble} Ultra Deep Field over $4.5$ arcmin$^2$ by Dunlop et al. (2016).

It is remarkable that an appreciable fraction of the individual, far-IR selected galaxies lie above the main sequence, i.e., at SFR values higher than expected on the basis of the average relationship at given $M_\star$. A common interpretation of these off main-sequence objects is that they are undergoing an episode of starburst triggered by a stochastic merger event.
Although these instances may well occur especially at low redshift $z\la 1$, our interpretation for the bulk of the objects at higher redshift $z\ga 1$ differs substantially. On the basis of the evolutionary track of individual galaxies, we envisage off main sequence objects to be caught in an early evolutionary stage, and still accumulating their stellar mass. Since the SFR changes slowly during the evolution, far-IR selection is unbiased with respect to the stellar mass; thus young starforming galaxies are found to be preferentially located above the main sequence, or better, to the left of it. As time goes by and stellar mass increases, the galaxy moves toward the average main sequence relationship, around which it will spend most of its lifetime. Afterwards, the SFR either slowly decreases because of gas exhaustion for galaxies with small final stellar mass, or is abruptly quenched by AGN feedback for galaxies with high final stellar mass. In particular, in the latter case the galaxy will then evolve passively to become a local early-type and will populate a region of the SFR vs. $M_\star$ diagram substantially below the main sequence. These loci of 'red and dead' galaxies are indeed observed locally (see Renzini \& Peng 2015), and start to be pinpointed even at high redshift (see Man et al. 2016).

Support to such an evolutionary picture comes from the estimates of the galaxy ages inferred from multiwavelength SED modeling by da Cunha et al. (2015) and Ma et al. (2015b). In Fig.~\ref{fig|MS_age_z2} we report the data points from the latter authors, with the galaxy age highlighted in different colors. It can be seen that galaxies at $z\sim 2$ (data points with black contours) located above the main sequence are preferentially younger and less massive, with ages substantially below a few $10^8$ yr; the ones more distant from the main sequence locus feature smaller and smaller ages.

In Fig.~\ref{fig|MS_UV_z2} we highlight that exploiting in Eq.~(\ref{eq|MS}) the UV-inferred SFR functions (dust corrected via the UV slope) originate a main sequence diagram in strong disagreement with the observations at high SFRs and/or stellar masses, because the SFR and stellar mass functions are considerably undersampled in the UV, especially at the high SFR/stellar mass end.

A more quantitative look at the number density of the main sequence outliers is presented in Fig.~\ref{fig|sSFR_dist}, where we show the distribution of specific SFR, i.e. sSFR$=\dot M_\star/M_\star$; it is obtained by slicing the main sequence diagram for different stellar mass bins (color-coded). The outcome when basing on the global (UV+far-IR) SFR functions (solid lines) is in remarkably good agreement with the observed distributions from Ilbert et al. (2015; see also Rodighiero et al. 2011; Schreiber et al. 2015); the latter is a $24\,\mu$m-selected sample extracted from the COSMOS and GOODS surveys, with robust SFR estimates via mid-/far-IR data from \textsl{Spitzer}/\textsl{Herschel}. On the other hand, the outcomes based on the UV-inferred SFR functions (dashed lines) substantially underpredict the observed number density of high sSFRs galaxies; this again reflects the undersampling of galaxies with high SFRs by the UV data.

A similar mismatch at high sSFR occurs for the semianalytic model by Wang et al. (2008b), reported in Fig.~\ref{fig|sSFR_dist} (dot-dashed lines) for the same stellar mass bins of the data. We also show the region (cyan shaded area) encompassed by the three recent merger-driven models considered in Guo et al. (2016). These refined models (see also Ciambur et al. 2013; Lamastra et al. 2013; Mitchell et al. 2014; Henriques et al. 2015; Schaye et al. 2015; Lacey et al. 2016), featuring both a quiescent and a starburst mode of star formation triggered by galaxy mergers, perform better than the Wang et al. model for stellar masses $M_\star\la$ a few $10^{10}\, M_\odot$, though still underpredicting somewhat the observed sSFR distribution at high sSFR. All in all, this comparison between merger-driven models and observations indicates that the merger-induced SFRs are too low and/or the duty cycles of the star formation episodes are too short (see also Fontanot et al. 2012; Brennan et al. 2016).

In  Fig.~\ref{fig|MS_evol} we confront the outcome of our computation for the average main sequence of starforming galaxies at different redshifts $z\sim 1$, $3$, and $6$ to the observational determinations by Speagle et al. (2014), finding good agreement. For comparison, the data points from individual observations presented in the previous Figures are also reported, with their estimated redshift highlighted in color. The observed redshift evolution of the main sequence is consistent with a scenario which traces the bulk of the star formation in galaxies back to local, in situ condensation processes. Specifically, at higher $z$ and in massive galaxies, the ISM is on average denser $\rho\propto (1+z)^3$; both the dynamical $t_d\propto 1/\sqrt{\rho}\propto (1+z)^{-3/2}$ and the cooling $t_c\propto 1/\rho\propto (1+z)^{-3} $ timescales becomes shorter, and the SFR $\dot M_\star\propto M_\star/\max[t_d,t_c]$ associated to a galaxy of given stellar mass is higher, so making the main sequence locus to shift upwards. We stress that moving toward higher redshift the fraction of off-main sequence objects decreases appreciably; this is because, given the evolution of the SFR function and the shorter age of the Universe, it is more and more difficult to spot galaxies of appreciably different ages and featuring very high SFRs.

\section{The main sequence of AGNs}\label{sec|AGN}

As discussed in Sect.~\ref{sec|intro}, nowadays it has been widely established that the mass of central BHs in massive, early-type galaxies is intimately linked to several properties of the hosts (e.g., velocity dispersion, stellar mass, etc.). However, a hot debate is still open on the physical origin of this link, in particular concerning the interplay between the star formation and BH accretion processes. Many studies pointed out the existence of an AGN main sequence, that relates the AGN luminosity to the SFR and/or stellar mass of the host starforming galaxy. We stress that the AGN main sequence is actually a coevolution plane, since involves both AGN and host galaxy properties; as such it differs from the stellar main sequence, which relates only star related variables.

In the following we discuss the BH accretion history as inferred from a wealth of multiwavelength observations. We show that it can be exploited to map the SFR functions into AGN bolometric luminosity functions, in excellent agreement with recent determinations. Then we turn to interpret and physically understand the AGN main sequence.

\subsection{BH accretion history}\label{sec|AGN_history}

The BH accretion history in high-$z$ starforming galaxies can be robustly constrained from a wealth of multiwavelength observations concerning: (i) the fraction of far-IR detected galaxies in X-ray selected AGNs (e.g., Mullaney et al. 2012b; Page et al. 2012; Rosario et al. 2012) and optically selected quasars (e.g., Mor et al. 2012; Wang et al. 2013b; Willott et al. 2015); (ii) the fraction of X-ray detected AGNs in far-IR/K-band selected host galaxies (e.g., Alexander et al. 2005; Mullaney et al. 2012a; Wang et al. 2013a; Johnson et al. 2013); (iii) stacking of undetected sources (e.g., Basu-Zych et al. 2013).

Lapi et al. (2014) and Aversa et al. (2015) have interpreted these data in terms of a very basic physical scenario of in situ coevolution. In a nutshell, this envisages that the early growth of the nuclear BH in high redshift starforming galaxies occurs under heavily dust enshrouded conditions. In the early stages $\tau\ll \tau_{\rm b}$, plenty of gas is available from the surroundings, and the BH accretes at substantial, mildly super-Eddington rates $\lambda\ga 1$, so as to develop radiatively-inefficient slim-disk conditions (see Abramowicz et al. 1988; Watarai et al. 2000; Li 2012; Madau et al. 2014; Volonteri et al. 2015).

During these early stages the BH luminosity is substantially smaller than that of the starforming host galaxy, that shines as a (sub-)mm bright source with an X-ray nucleus. After a time $\tau\approx \tau_{\rm b}\la$ Gyr, the nuclear power progressively increases to values similar or even exceeding that of the host galaxy. Strong BH winds remove interstellar gas and dust while quenching star formation, so that the system behaves as an optical quasar. Residual gas present in the central regions of the galaxy can be accreted onto the BH at progressively lower, sub-Eddington accretion rates; the accretion disk becomes thin, yielding the standard SEDs observed in type-1 AGNs (Elvis et al. 1994; Hao et al. 2014). In quantitative terms, the typical AGN lightcurve can be described as (Yu \& Lu 2004; Lapi et al. 2014; Aversa et al. 2015)
\begin{eqnarray}\label{eq|AGNlightcurve}
L_{\rm AGN}(\tau|M_{{\rm BH}},t)\nonumber &=& L_{{\rm AGN},{\rm b}}\,e^{(\tau -\tau_{\rm b})/\tau_{\rm ef}} ~~~~ 0 \le \tau \le\tau_{\rm b}\\
\\
\nonumber &=& L_{{\rm AGN},{\rm b}}\, e^{-(\tau-\tau_{\rm b})/\tau_{\rm AGN}} ~~~~ \tau\ge \tau_{\rm b}~. \end{eqnarray}
During the early phase up to the time $\tau_{\rm b}$
the BH mass increases exponentially with characteristic timescale $\tau_{\rm ef}$ up to a value $M_{{\rm BH}, {\rm b}}$; the AGN emits with an Eddington ratio $\lambda\ga 1$ until reaching a peak luminosity $L_{{\rm AGN},{\rm b}}=\lambda\,M_{{\rm BH}, \rm {\rm b}}\, c^2/ t_{\rm Edd}$. Then a late phase when the luminosity exponentially declines with a characteristic timescale $\tau_{\rm AGN}$ follows. The quantities $\lambda$ and $\epsilon$ refer to the average radiative efficiency and Eddington ratio during the early phase, so that the $e-$folding time reads $\tau_{\rm ef}=\epsilon\, t_{\rm Edd}/\lambda\,(1-\epsilon)$.

As to the characteristic time $\tau_{\rm AGN}$ of the declining phase, the data on the fraction of starforming host galaxies in optically-selected quasars suggest a value $\tau_{\rm AGN}\approx 3\, \tau_{\rm ef}$ for AGNs with peak luminosities $L_{{\rm AGN},\rm b}\ga 10^{13}\, L_\odot$; on the other hand, AGNs with lower peak luminosity are constrained to fade more drastically after the peak (Lapi et al. 2014 and Aversa et al. 2015). These behaviors can be smoothly rendered as
\begin{equation}\label{eq|tau_d}
\tau_{\rm AGN} = 3\, \tau_{\rm ef}\, \left[1-{1\over 2}\,{\rm
erfc}\left({1\over 2}\,\log{L_{{\rm AGN},\rm b}\over 10^{13} L_\odot}\right)\right]~.
\end{equation}

As to the Eddington ratio $\lambda$, its evolution with redshift is strongly constrained by various observations, including the bright end of the optical AGN luminosity function, the observed Eddington ratio function (see Vestergaard \& Osmer 2009; Kelly \& Shen 2013; Schulze et al. 2015), and the fraction of AGN hosts with given stellar mass as a function of the Eddington ratio (see Aird et al. 2012; Bongiorno et al. 2012). The implied dependence of $\lambda$ on redshift $z$ can be rendered approximately as (see Lapi et al. 2014; Aversa et al. 2015)
\begin{equation}\label{eq|lambda_z}
\lambda(z)= 4\,\left[1-{1\over 2}\,{\rm erfc}\left({z-2\over 3}\right)\right]~.
\end{equation}
Note that during the early stages $\lambda\ga \lambda_{\rm thin}\approx 0.3$ holds and a radiatively-inefficient slim accretion disc is expected (Laor \& Netzer 1989); since the Eddington ratio lowers rapidly, a thin accretion disc develops during the late evolution.

As to the radiative efficiency $\epsilon$, a simple prescription relates it and the Eddington ratio $\lambda$ for slim/thin discs (see Abramowicz et al. 1988; Watarai et al. 2000; Blandford \& Begelman 2004; Li 2012; Madau et al. 2014) in the form
\begin{equation}\label{eq|epslambda}
\epsilon = {\epsilon_{\rm thin}\over 2}\, {\lambda\over e^{\lambda/2}-1}~;
\end{equation}
here $\epsilon_{\rm thin}\sim 0.057-0.32$ refers to the efficiency during the thin disc phase, that depends on the BH spin (Thorne 1974). A fiducial value $\epsilon_{\rm thin}=0.1$ is suggested by statistical investigations based on the continuity equation (e.g., Cao 2010; Aversa et al. 2015) and from observations of individual systems (see Davis \& Laor 2011; Raimundo et al. 2012; Wu et al. 2013). In the early stages when mildly super-Eddington accretion occurs with $\lambda\ga$ a few, the radiative efficiency takes on small values $\epsilon\la 0.3\, \epsilon_{\rm thin}\approx 0.03$, while in the late stages when sub-Eddington accretion occurs, it quickly approaches the thin disc value $\epsilon\approx \epsilon_{\rm thin}\approx 0.1$.

The final BH mass $M_{\rm BH}$ is simply written in terms of the peak mass $M_{{\rm BH},{\rm b}}$ appearing in Eq.~(\ref{eq|AGNlightcurve}) via $L_{\rm AGN,b}$; it reads
\begin{eqnarray}\label{eq|BH_mass}
\nonumber M_{{\rm BH}} &=& \int_0^{\infty}\, d\tau' {1-\epsilon\over \epsilon\, c^2}\, L_{\rm AGN}(\tau') = \\
\\
\nonumber &=& M_{{\rm BH},{\rm b}}\,\left[1+f_\epsilon\,{\tau_{\rm AGN}\over \tau_{\rm ef}}\right]~.
\end{eqnarray}
The factor $f_\epsilon$ takes into account the modest change of
$(1-\epsilon)/\epsilon$ during the declining phase, when $L_{\rm AGN}(\tau)$ and $\lambda(\tau)$ decrease almost exponentially and $\epsilon(\tau)$ increases according to Eq.~(\ref{eq|epslambda}); a value $f_\epsilon\approx 0.8$ applies to a good accuracy.

The evolution of the AGN luminosity and BH  mass during the galaxy timetime $\tau$ is illustrated in Fig.~\ref{fig|lightcurves}.

\subsection{Mapping the SFR functions into the AGN luminosity functions}\label{sec|AGN_LF}

We stress that the above BH accretion history and timescales not only have been inferred from a plethora of observational data, but also have been validated in Aversa et al. (2015) by showing that the BH mass function obtained via the continuity equation approach well reproduces the current observational constraints at different redshifts (e.g., Vika et al. 2009; Shankar et al. 2009, 2013; Willott et al. 2010; Li et al. 2011; Ueda et al. 2014).

Here we provide a further validation, by examining the AGN statistics from a galaxy evolution viewpoint, basing on the SFR functions and on a deterministic BH accretion history consistent with the Eddington ratio distribution at various redshifts (cf. Figs.~8 and 9 in Aversa et al. 2015).
This is similar in spirit with, but different operationally from, the approach by Caplar et al. (2015), who recovered the AGN luminosity functions from the stellar mass functions and from an ad hoc Eddington ratio distribution.

To relate the BH accretion and star formation histories, we set the final BH/stellar mass ratio $M_{\rm BH}/M_\star$ to average values $\approx 10^{-3}$ with a scatter $\approx 0.4$ dex, as directly measured in the local Universe (see Shankar et al. 2016). Direct observational determinations and statistical estimates based on the abundance matching between the stellar and BH mass functions out to $z\sim 2$ indicate model-independently, though with large uncertainties, that the $M_{\rm BH}/M_\star$ ratio weakly increases with redshift (e.g., Haring \& Rix 2004; Peng 2007; Jahnke \& Macci\'o 2011; Kormendy \& Ho 2013; Aversa et al. 2015; Shankar et al. 2016). We render this average evolutionary behavior as
\begin{equation}\label{eq|BH_Mstar_ratio}
{M_{\rm BH}\over M_\star}= 1.5\times 10^{-3}\, \left({1+z\over 2}\right)^{\zeta}~,
\end{equation}
with a fiducial value $\zeta\approx 1/2$, and assume a scatter of $0.4$ dex independent of redshift around this mean relationship. We checked that our results on the AGN luminosity functions and AGN main sequence are marginally affected by varying $\zeta$ in the plausible range from $0$ to $1$; note that a decrease in $\zeta$ is easily offset by a moderate increase in the scatter toward high redshift, and viceversa.

The relative time spent by the AGN in a given logarithmic bin of luminosity $L_{\rm AGN}$, i.e. the AGN duty cycle, just reads
\begin{equation}\label{eq|AGN_duty}
{{\rm d}\delta\over {\rm d}\log L_{\rm AGN}}={\tau_{\rm ef}+\tau_{\rm AGN}\over \tau_{\rm b}+\xi\, \tau_{\rm AGN}}\,\ln 10~;
\end{equation}
here the total duration of the AGN luminous phase is taken as $\tau_{\rm b}+\xi\, \tau_{\rm AGN}$ with $\xi\approx 3$, since after that time the luminosity is already decreased by a factor $e^{-\xi}\la 0.05$
and the BH mass has already attained its final value to a very good approximation; anyway, variations of this parameter do not appreciably affect our results.

Then the AGN bolometric luminosity function can be straightforwardly computed as
\begin{equation}\label{eq|AGN_LF}
{{\rm d} N\over {\rm d}\log L_{\rm AGN}} = \int{{\rm d}\log\dot M_\star}\,{{\rm d} N\over {\rm d}\log\dot M_\star}\,{{\rm d}\delta\over {\rm d}\log L_{\rm AGN}}~.
\end{equation}
The outcome is illustrated in Fig.~\ref{fig|AGN_LF} at various redshifts, and pleasingly agrees with the most recent observational determinations. This further validates the proposed BH accretion history.

Basing on the UV-inferred SFR functions would clearly undersample the AGN luminosity functions at the bright end, since the number of galaxies with high SFR and high stellar mass, hence with high BH mass and AGN luminosity, is itself underestimated. Note that, as shown by the dotted lines in Fig.~\ref{fig|SFR_func}, AGNs with appreciable X-ray luminosity $L_X\ga 10^{42}$ erg s$^{-1}$ (bolometric corrections by Hopkins et al. 2007 have been adopted) are hosted by galaxies with SFRs $\dot M_\star\ga 100\, M_\odot$ yr$^{-1}$, but their number density is smaller than that of the overall starforming population by a factor $\la 10^{-1}$; this is because of the duty cycle (cf. Eq.~\ref{eq|AGN_duty}), which reflects the nearly constant behavior of the SFR vs. the exponential growth of the BH accretion rate during the galaxy lifetime.

In Fig.~\ref{fig|cosm} we report the cosmological evolution of the BH accretion rate density, computed as
\begin{equation}
\rho_{\dot M_{\rm BH}} = \int{\rm d}\log L_{\rm AGN}\, {{\rm d}N\over {\rm d}\log L_{\rm AGN}}\, {L_{\rm AGN}\over \epsilon\, c^2}
\end{equation}
with a reference value $\epsilon\approx 0.1$ of the radiative efficiency.
We show the results for two minimum values of $L_{X,\rm min}\ga 10^{42}$ and $10^{44}$ erg s$^{-1}$, to cope with the observational limits in the estimate of the luminosity functions at $z\la 3$ and $z\ga 3$, respectively. Our combined results well compare with the observational estimates by Aird et al. (2015) from X-ray selection and by Delvecchio et al. (2014) from an IR perspective.  It is interesting to note that the cosmic BH accretion rate and SFR densities feature a similar evolution, with a peak around $z\sim 2-3$ and a decline toward higher redshift. We stress that this is a statistical outcome originated by the shape of the SFR and AGN luminosity functions; by no means it implies that within an individual galaxy the BH accretion rate is \emph{always} proportional to the SFR.

\subsection{Interpreting the AGN main sequence}\label{sec|AGN_MS}

Given consistent SFR functions, AGN duty cycles and luminosity functions, we can investigate the AGN main sequence.  We start from discussing the $L_{\rm AGN}$ vs. SFR diagram. The number of objects per logarithmic bins of AGN luminosity and SFR is given by
\begin{equation}\label{eq|AGN_MS}
{{\rm d^2} N\over {\rm d}\log \dot M_\star\,{\rm d}\log L_{\rm AGN}} \simeq {{\rm d} N\over {\rm d}\log\dot M_\star}\,{{\rm d}\delta\over {\rm d}\log L_{\rm AGN}}~.
\end{equation}
The outcome at the representative redshift $z\approx 2$ is presented in Fig.~\ref{fig|CP_z2}, where the color-code indicates the logarithmic number density of objects per unit comoving volume (in Mpc$^{-3}$). The lilac line with errorbars illustrates the the number density-weighted mean relationship $\langle\dot M_\star\rangle$ at given $L_{\rm AGN}$ with its $2\sigma$ variance when adopting, for a fair comparison with data, a SFR detection limit of $10^2\, M_\odot$ yr$^{-1}$. Our deterministic evolutionary tracks for the star formation and corresponding BH accretion history in individual galaxies (dotted lines), and the locus where the AGN and SFR luminosities match (dashed line), are also illustrated.

For AGN luminosities $L_{\rm AGN}<L_{\rm SFR}$ lower than those associated with star formation, the tracks of individual objects and the mean relationships are flat; in fact, this occurs for galactic ages $\tau\la \tau_{\rm b}$ when the SFR is roughly constant while the AGN luminosity grows exponentially. Then on moving toward higher AGN luminosities
$L_{\rm AGN}\ga L_{\rm SFR}$ that are attained for galactic ages $\tau\ga \tau_{\rm b}$, the SFR gets rapidly suppressed and the AGN luminosity fades, so that the evolutionary tracks of individual objects move toward the bottom left region of the diagram. Contrariwise, the mean relationship starts to increase, being statistically dominated by objects with higher and higher SFR; this is because to achieve a higher AGN luminosity, the BH must reside in a galaxy with larger SFR.

\begin{figure}
\epsscale{0.8}\plotone{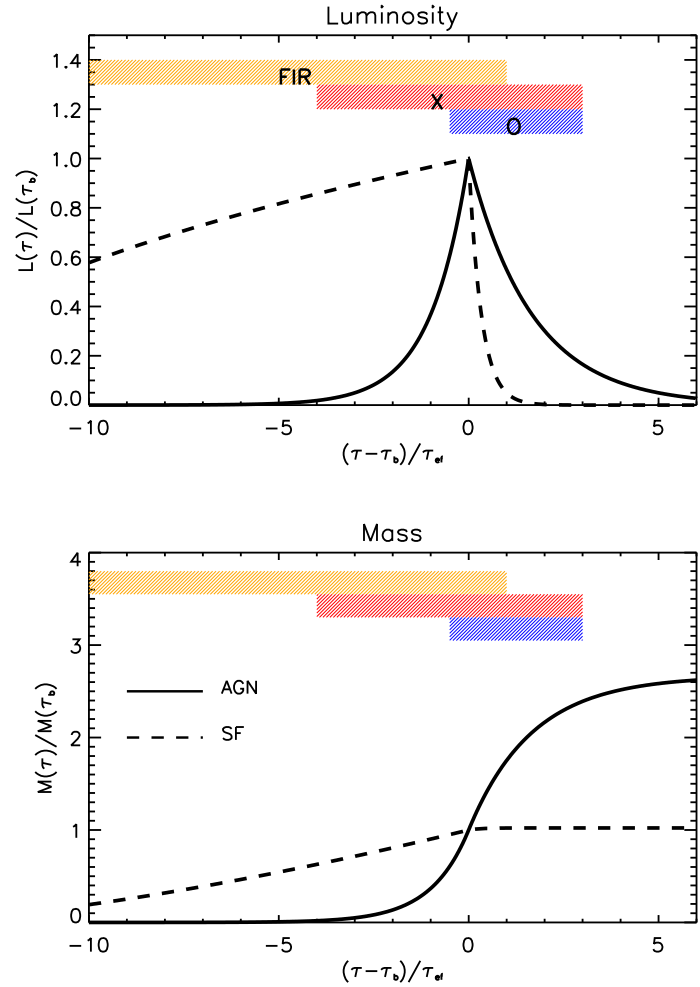}\caption{Evolution with galactic age (in units of BH $e-$folding time $\tau_{\rm ef}$) of the luminosity (top panel) and mass (bottom panel), normalized at the time $\tau_{\rm b}$ when the AGN luminosity peaks and the star formation is quenched by the AGN feedback. Solid lines refer to AGN-related quantities and dashed lines to star formation-related quantities. The orange area sketches the stage when the starforming galaxy is dust-enshrouded and appears as a far-IR bright source; the red area sketches the stage when the AGN X-ray (intrinsic) luminosity overwhelms that associated to star formation; the blue area sketches the optical phase, setting in when the quasar feedback removes gas and dust from the medium and quenches star formation in the host.}\label{fig|lightcurves}
\end{figure}

\begin{figure}
\epsscale{1}\plotone{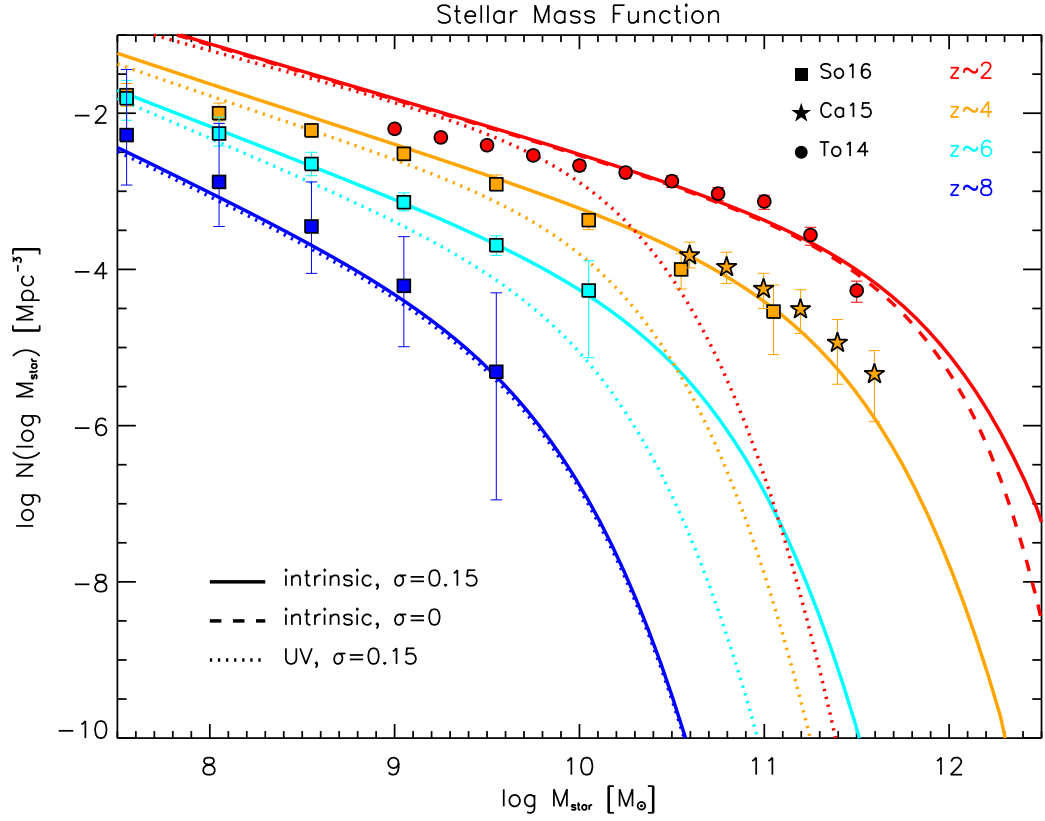}\caption{The stellar mass function of starforming galaxies at $z=2$ (red), $4$ (orange), $6$ (cyan), and $8$ (blue) as derived from the SFR functions (cf. Fig.~\ref{fig|SFR_func}) and from the star formation timescale $\tau_{\rm b}$ (cf. Eq.~\ref{eq|tau_burst}) via the continuity equation (cf. Eq.~\ref{eq|continuity}). Solid lines refer to the global (UV+far-IR) SFR functions with a scatter $\sigma=0.15$ dex in the starformation timescale $\tau_{\rm b}$, dashed line (only at $z=2$ for clarity) refers to zero scatter, and dotted lines to the (dust-corrected) UV-inferred SFR functions. Data are from Tomczak et al. (2014; circles), Caputi et al. (2015; stars) and Song et al. (2016; squares).}\label{fig|STAR_MF}
\end{figure}

\begin{figure}
\epsscale{1}\plotone{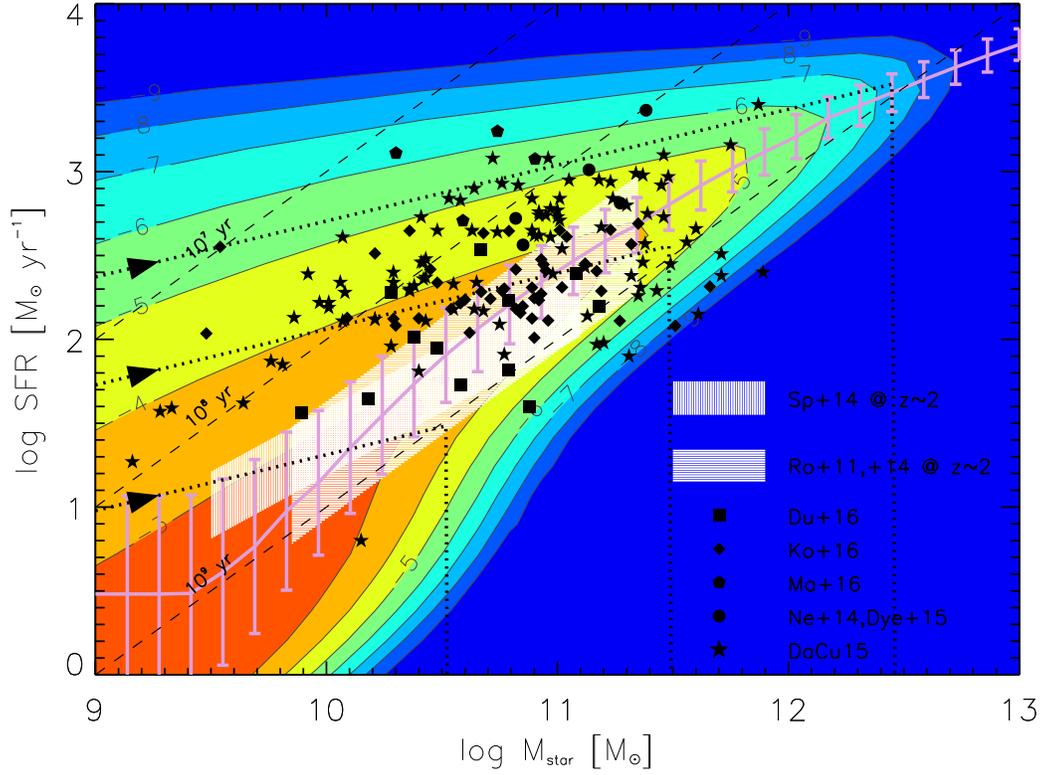}\caption{The main sequence of starforming galaxies at $z\approx 2$, based on the global (UV+far-IR) SFR functions. Colored contours illustrate the number density of galaxies (labels are in log units of Mpc$^{-3}$) in the SFR vs. $M_\star$ plane. The lilac line with errorbars illustrates the mean relationship with its $2\sigma$ scatter. The dotted lines show three evolutionary tracks (forward time direction indicated by arrows) for galaxies with a given final stellar mass of about $10^{10.5}$, $10^{11.5}$, $10^{12.5}\, M_\odot$. The dashed lines show the timescale $M_\star/\dot M_\star=10^7$, $10^8$ and $10^9$ yr as labeled. The white shaded areas are the observational determinations of the main sequence (based on statistics of large samples) by Rodighiero et al. (2011, 2014; horizontal line pattern), and by Speagle et al. (2014; vertical line pattern). Filled black symbols (error bars omitted for clarity) refer to far-IR data for individual objects by Dunlop et al. (2016; squares), Koprowski et al. (2016; diamonds), Ma et al. (2015b; pentagons), Negrello et al. (2014) plus Dye et al. (2015; circles), and Da Cunha et al. (2015; stars).}\label{fig|MS_z2}
\end{figure}

\begin{figure}
\epsscale{1}\plotone{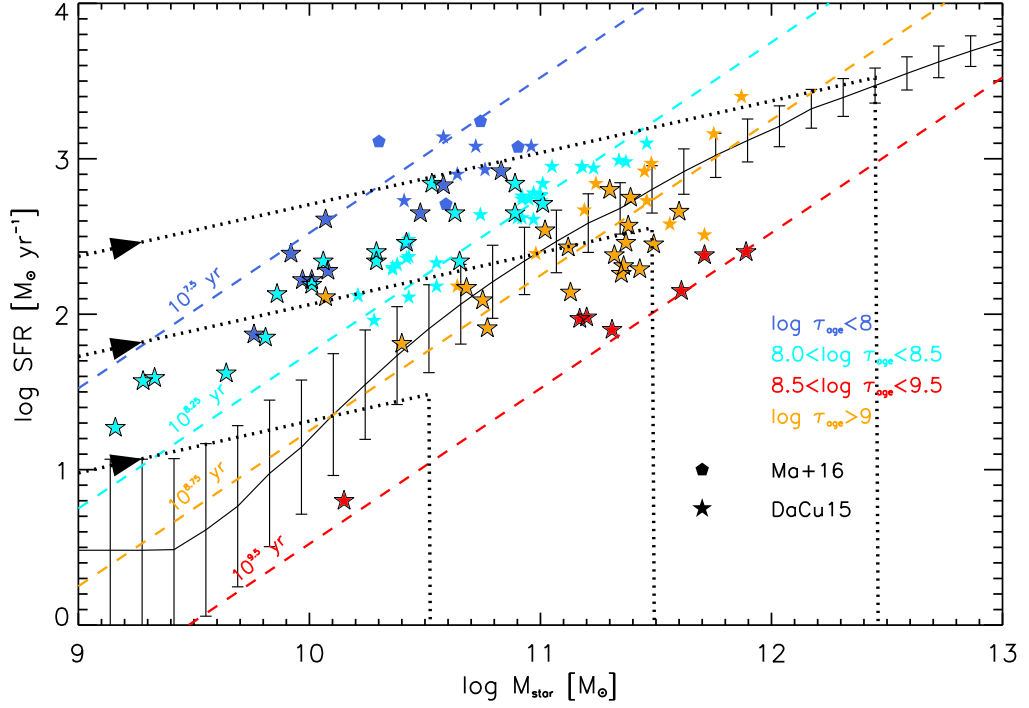}\caption{The main sequence of starforming galaxies at $z\approx 2$, based on the global (UV+far-IR) SFR functions. The solid line with errorbars illustrates the mean relationships with its $2\sigma$ scatter. The dotted lines show three evolutionary tracks (forward time direction indicated by arrows) for galaxies with a given final stellar mass of about $10^{10.5}$, $10^{11.5}$, $10^{12.5}\, M_\odot$.  The dashed lines shows the timescale $M_\star/\dot M_\star=10^{7.5}$, $10^{8.25}$, $10^{8.75}$, and $10^{9.5}$ yr as labeled. Filled symbols (error bars omitted for clarity) refer to far-IR data for individual objects by Ma et al. (2015b; pentagons) and Da Cunha et al. (2015; stars), color-coded according to the SED-inferred age $\tau_{\rm age}$ in bins $\tau_{\rm age}\la 10^8$ (blue), $10^8\la \tau_{\rm age}\la 10^{8.5}$ (cyan), $10^{8.5}\la \tau_{\rm age}\la 10^{9}$ (orange), and $\tau_{\rm age}\ga 10^{9}$ (red). Data points for galaxies at $1\la z\la 3$ are highlighted by a black contour.}\label{fig|MS_age_z2}
\end{figure}

\begin{figure}
\epsscale{1}\plotone{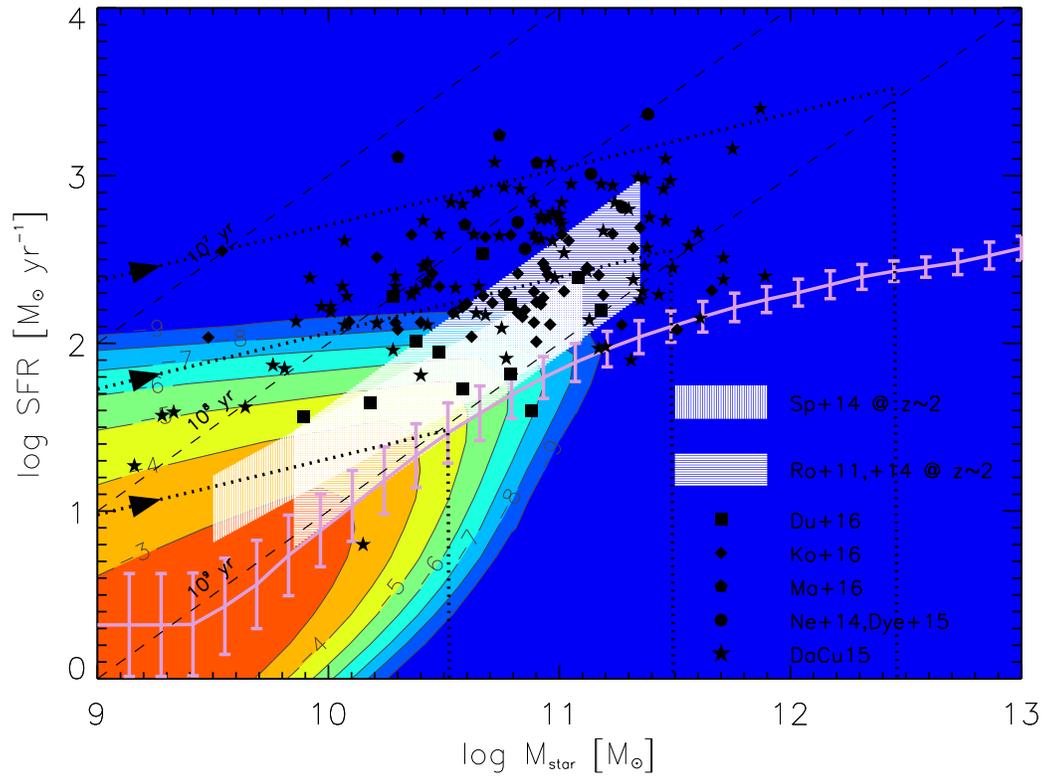}\caption{Same as Fig.~\ref{fig|MS_z2}, but based on the (dust-corrected) UV-inferred SFR function.}\label{fig|MS_UV_z2}
\end{figure}

\begin{figure}
\epsscale{1}\plotone{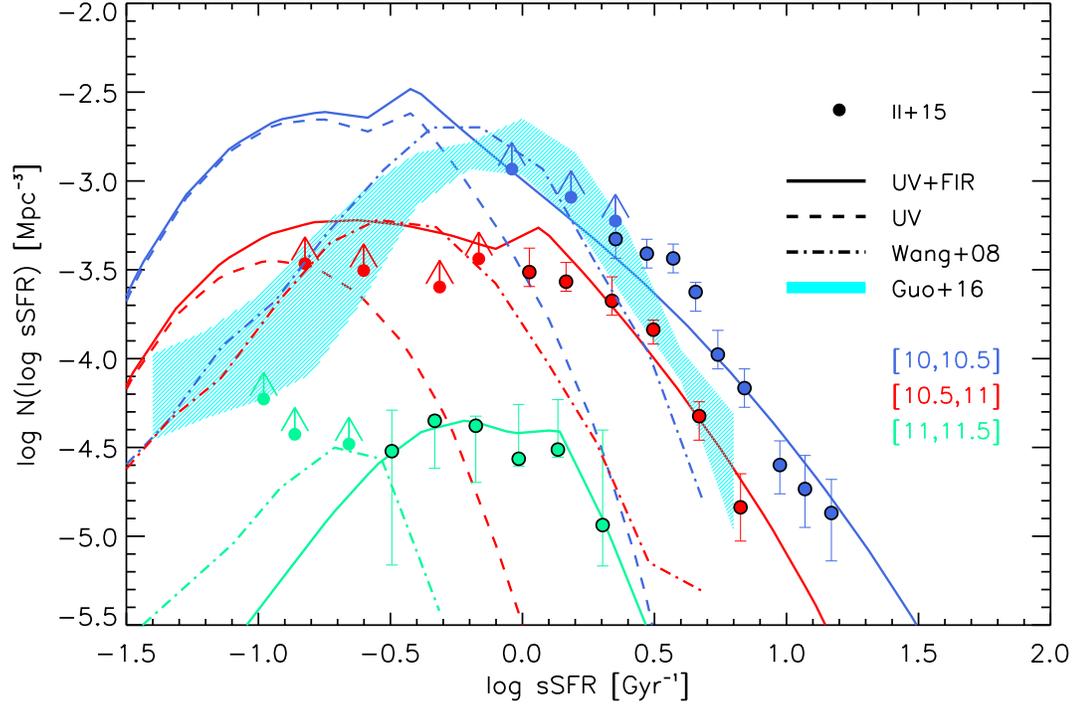}\caption{The distribution of sSFR$=\dot M_\star/M_\star$ at $z\sim 2$. Data (filled circles) are from Ilbert et al. (2015), with colors referring to different stellar mass bins: $\log M_\star/M_\odot\in [10,10.5]$ is coded in blue, $[10.5,11]$ in red and $[11,11.5]$ in green. Solid lines illustrate the outcomes when using the global (UV+far-IR) SFR functions, while dashed lines are when using the (dust-corrected) UV-inferred SFR functions. Dot-dashed lines illustrate the prediction of the merger-driven semianalytic model by Wang et al. (2008b, as reported by Ilbert et al. 2015) for the same stellar mass bins of the data. The cyan shaded area illustrates the region encompassed by the three recent merger-driven models considered in Guo et al. (2016) for the lowest stellar mass bin.}\label{fig|sSFR_dist}
\end{figure}

\begin{figure}
\epsscale{1}\plotone{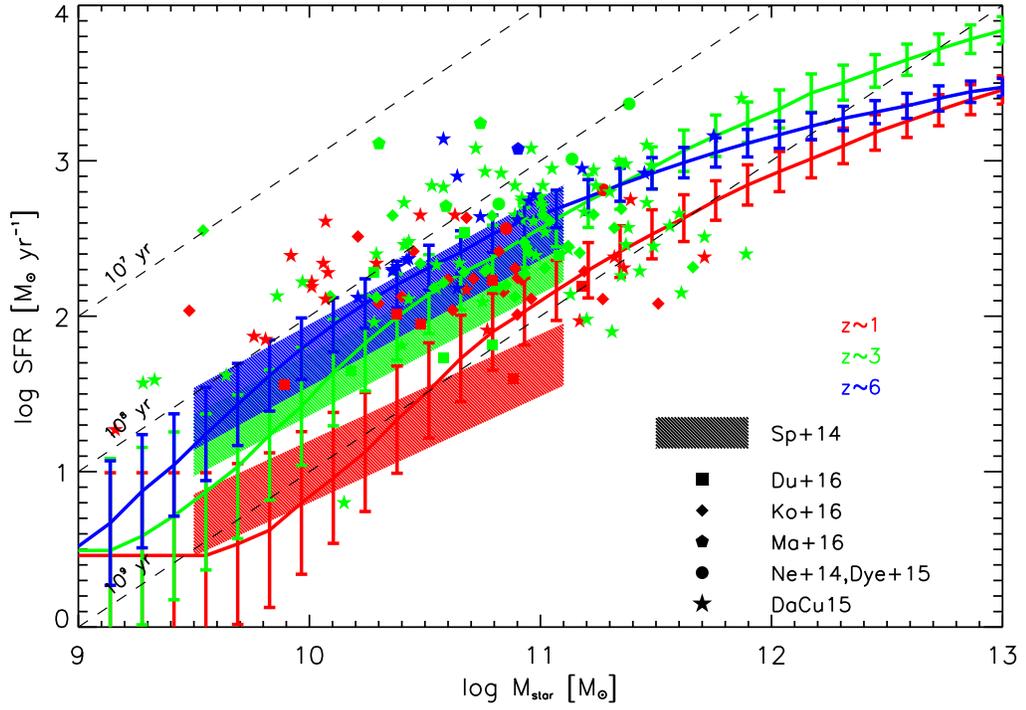}\caption{The main sequence of starforming galaxies at $z\approx 1$ (red), $3$ (green), and $6$ (blue), based on the global (UV+far-IR) SFR functions. The solid lines with errorbars illustrate the mean relationships with its $2\sigma$ scatter. The dashed lines shows the timescale $M_\star/\dot M_\star=10^7$, $10^8$ and $10^9$ yr as labeled. The shaded areas are the observational determinations of the main sequence at different redshifts (based on statistics of large samples) by Speagle et al. (2014). Filled symbols (error bars omitted for clarity) refer to far-IR data for individual objects by Dunlop et al. (2016; squares), Koprowski et al. (2016; diamonds), Ma et al. (2015b; pentagons), Negrello et al. (2014) plus Dye et al. (2015; circles), and Da Cunha et al. (2015; stars), color-coded according to redshift bins $z\la 2$ (red), $2\la z\la 4$ (green), and $z\ga 4$ (blue).}\label{fig|MS_evol}
\end{figure}

\begin{figure}
\epsscale{1}\plotone{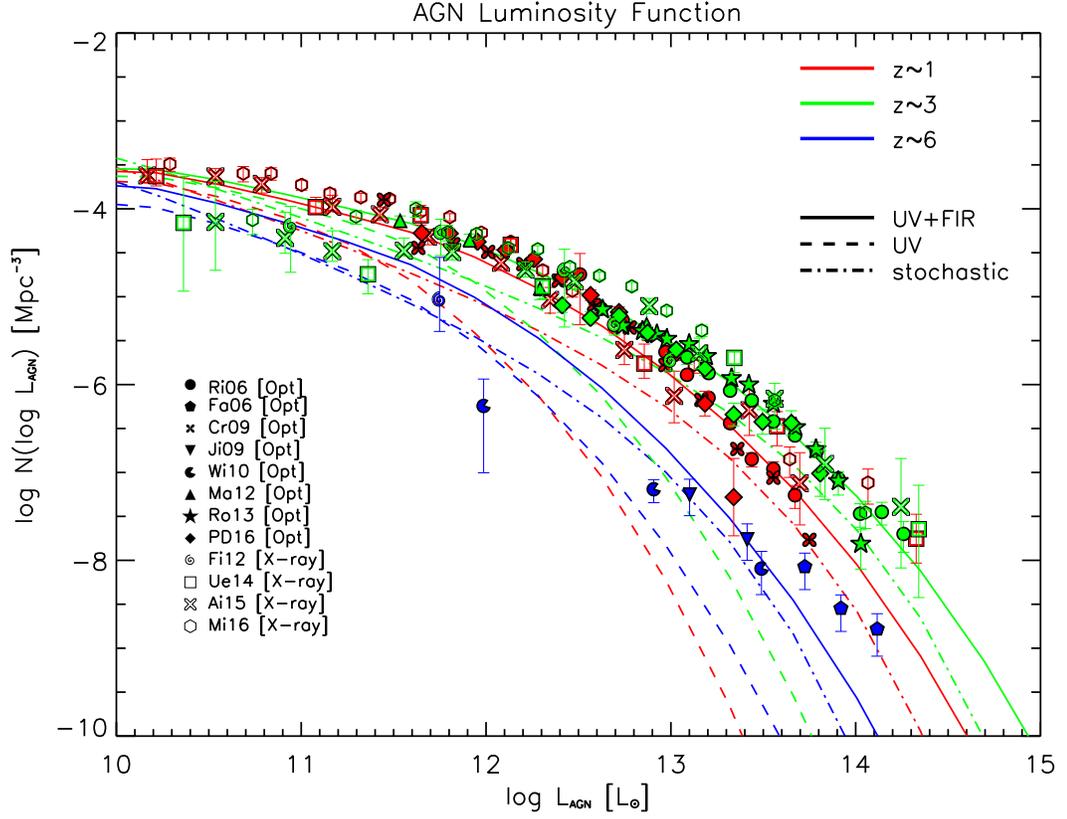}\caption{The (bolometric) AGN luminosity functions at redshifts $z=1$ (red lines), $3$ (green lines) and $6$ (blue lines), as reconstructed from the SFR functions and the AGN duty cycle associated to the deterministic BH accretion history of \S~\ref{sec|AGN_history}. Solid lines refer to our outcomes based on the global (UV+far-IR) SFR functions, while  dashed lines refer to the (dust corrected) UV-inferred SFR functions. Dot-dashed lines are obtained by adopting the stochastic variability model by Hickox et al. (2014) and Stanley et al. (2015), inspired by the merger-driven scenario (see Eqs.~\ref{eq|hickox_dist} and \ref{eq|hickox_ave}). Optical data (filled symbols) are from Richards et al. (2006; circles), Fan et al. (2006; pentagons), Croom et al. (2009; crosses), Jiang et al. (2009; inverse triangles), Willott et al. (2010; pacmans), Masters et al. (2012; triangles), Ross et al. (2013; stars), and Palanque-Delabrouille et al. (2016; diamonds); X-ray data (empty symbols) are from Fiore et al. (2012; spirals), Ueda et al. (2014; squares), Aird et al. (2015; big cross), and Miyaji et al. (2015; hexagons). The X-ray and optical luminosities have been converted to bolometric by using the corrections from Hopkins et al. (2007; see their Figure 1), while the number densities have been corrected for obscured AGNs following Ueda et al. (2003, 2014).}\label{fig|AGN_LF}
\end{figure}

\begin{figure}
\epsscale{1}\plotone{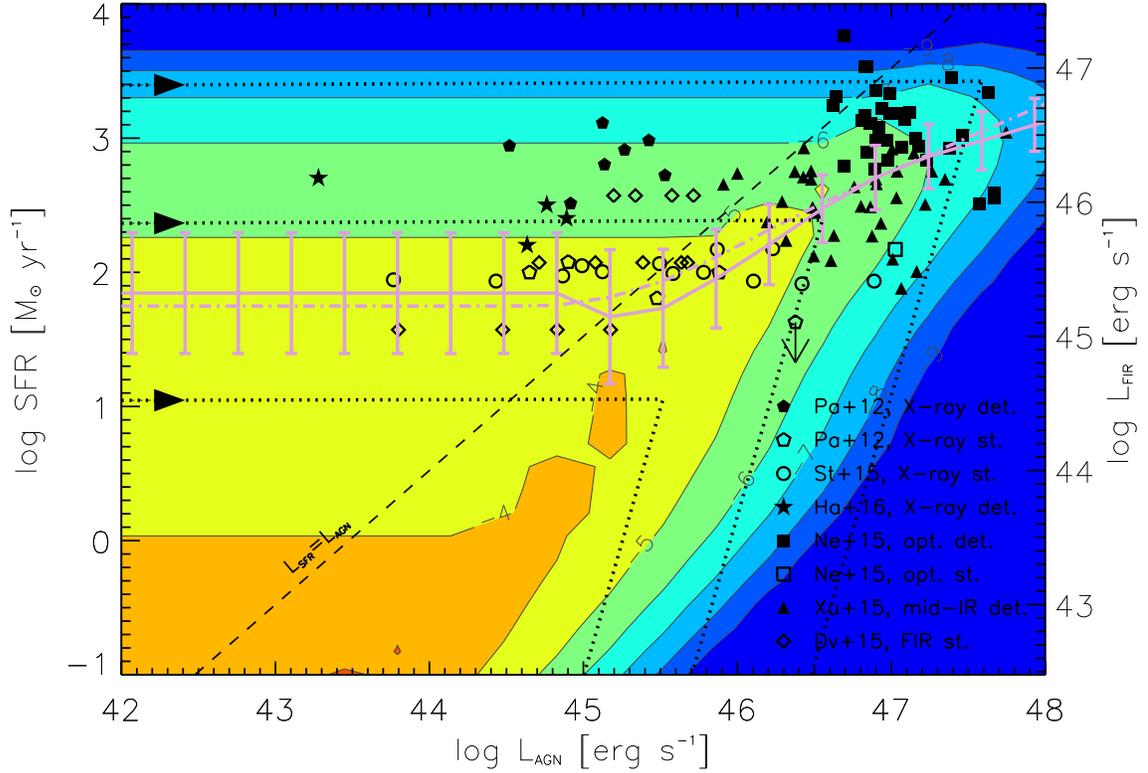}\caption{Relationship between the SFR (left axis) or the far-IR luminosity (right axis) vs. the (bolometric) AGN luminosity at $z\approx 2$, based on the global (UV+far-IR) SFR functions and the AGN duty cycle associated to the deterministic BH accretion history of \S~\ref{sec|AGN_history}. Colored contours illustrate the number density of galaxies plus AGNs (labels are in log units of Mpc$^{-3}$). The lilac line with errorbars illustrate the mean relationship with its $2\sigma$ scatter. The dashed line represents the locus where $L_{\rm SFR}=L_{\rm AGN}$. The dotted lines show three evolutionary tracks (forward time direction indicated by arrows) for objects with a given peak AGN luminosity of $10^{45.5}$, $10^{46.5}$, $10^{47.5}$ erg s$^{-1}$. Dot-dashed line is the average relationship obtained by adopting the variability model by Hickox et al. (2014) and Stanley et al. (2015), inspired by the merger-driven scenario (see Eqs.~\ref{eq|hickox_dist} and \ref{eq|hickox_ave}). Data are from: Page et al. (2012; pentagons) for both individual (filled symbols) and stacked (open symbols) X-ray selected sources; Stanley et al. (2015; circles) for stacked X-ray selected sources; Harrison et al. (2016; stars) for individual X-ray selected sources; Netzer et al. (2016; squares) for both individual and stacked optically selected sources; Xu et al. (2015; triangles) for individual mid-IR selected sources; Delvecchio et al. (2015; diamonds) for stacked far-IR selected sources.}\label{fig|CP_z2}
\end{figure}

In the diagram we illustrate for comparison data points from individual and stacked observations of X-ray selected AGNs (Page et al. 2012; Stanley et al. 2015; Harrison et al. 2016), mid-IR selected AGNs (Xu et al. 2015), and optically selected quasars (Netzer et al. 2016). The position on the diagram of these data points can be easily understood as a selection effect (cf. Fig.~\ref{fig|lightcurves}). Optical selection tends to pick up objects close to the peak of AGN luminosity when in the host dust has been partially removed and the SFR starts to be quenched; X-ray selection can pick up objects before or after the AGN peak, hence with SFR in the host still sustained or suppressed, respectively. Mid-IR selection with the current observational limits strikes an intermediate course between the former two.

For the sake of completeness, we also report data for a sample of far-IR selected galaxies, where the bolometric AGN luminosity has been estimated by stacking of X-ray fluxes (Delvecchio et al. 2015). In each of the SFR bins, the nuclear luminosity span a range of values, because the galaxy can be picked up at anytime before the peak, when the AGN luminosity can have vastly different values. Note that the contour levels reported in the figure apply if the primary selection is in AGN luminosity, and not in far-IR emission associated to star formation in the host.

In Fig.~\ref{fig|CP_z2} the average result based on the variability model by Hickox et al. (2014) and Stanley et al. (2015) is also shown. These authors adopt an Eddington ratio (or AGN luminosity) distribution inspired by merger-driven models of galaxy formation, that translates into a duty cycle with shape
\begin{equation}\label{eq|hickox_dist}
{{\rm d}\delta \over {\rm d}\log L_{\rm AGN}}\propto \left(L_{\rm AGN}\over 100\,\langle L_{\rm AGN}\rangle\right)^{-\omega}\, e^{-L_{\rm AGN}/100\,\langle L_{\rm AGN}\rangle}~,
\end{equation}
where $\omega\approx 0.2$ (actually Hickox et al. 2004 adopt $\omega\approx 0.6$, but the results change little); the distribution is normalized to unity over the range $L_{\rm AGN}>10^{-2}\,\langle L_{\rm AGN}\rangle$.

The above authors also assume a constant ratio $\dot M_{\rm BH}/\dot M_\star \approx 1/3000$ between the BH accretion rate and the SFR, as observed in the local Universe (e.g., Chen et al. 2013); this provides a link between SFR and average AGN luminosity $\langle L_{\rm AGN}\rangle$ in the form
\begin{equation}\label{eq|hickox_ave}
\langle L_{\rm AGN}\rangle \approx 2\times 10^{42}\, {\rm erg~s^{-1}}\times {\dot M_\star\over M_\odot~{\rm yr^{-1}}}
\end{equation}
Such a stochastic model produces a result (lilac dot dashed line) similar to our (lilac solid line) on the SFR vs. $L_{\rm AGN}$ diagram. However, as shown from Fig.~\ref{fig|AGN_LF} it considerably underpredicts the observed bolometric AGN luminosity functions at $z\ga 1$, especially at the bright end. This is because Eq.~(\ref{eq|hickox_ave}) implies a too low normalization of the SFR to the BH accretion rate ($1/3000$) at high redshift, and Eq.~(\ref{eq|hickox_dist}) is not sufficiently broad at the bright end (cf. Bongiorno et al. 2012; Nobuta et al. 2012; Kelly \& Shen 2013).

\begin{figure}
\epsscale{1}\plotone{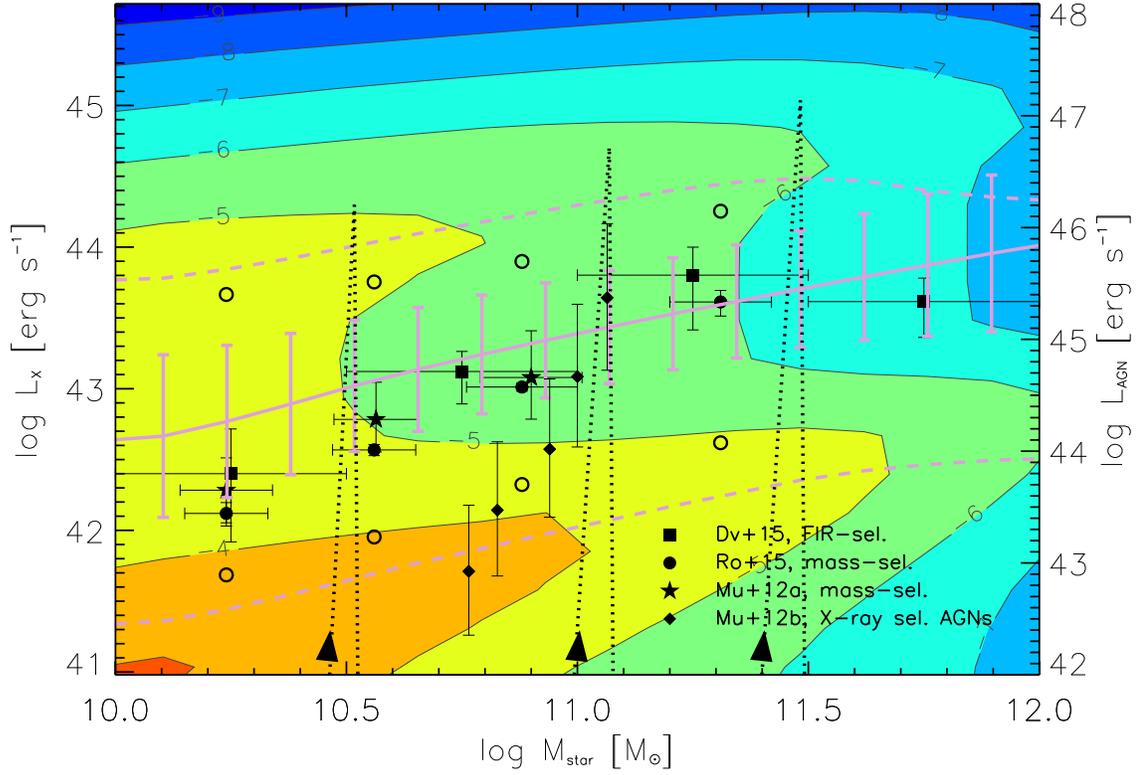}\caption{Relationship between the X-ray luminosity (left axis) or the bolometric AGN luminosity (right axis, strictly valid only for detected sources) vs. the host galaxy stellar mass at $z\approx 2$. The relationship is based on the global (UV+far-IR) SFR functions and the AGN duty cycle associated to the deterministic BH accretion history of \S~\ref{sec|AGN_history}. Colored contours illustrate the number density of objects (labels are in log units of Mpc$^{-3}$). The lilac line with errorbars illustrates the relationship averaged over detected and stacked sources (dashed lines) with its $2\sigma$ variance, for a detection limit at $L_X\approx 10^{43}$ erg s$^{-1}$. The dotted lines show three evolutionary tracks (forward time direction indicated by arrows) for objects with a given peak AGN luminosity of $10^{46}$, $10^{46.5}$, $10^{47}$ erg s$^{-1}$. Data are from Delvecchio et al. (2015; squares) for far-IR selected sources, from Mullaney et al. (2015b; diamonds) for X-ray selected sources, and from Mullaney et al. (2012a; stars) and Rodighiero et al. (2015; circles) for mass-selected samples; the open circles illustrate the average for detected (top) and stacked (bottom) sources in the Rodighiero et al. (2015) data.}\label{fig|CPmass_z2}
\end{figure}

We now turn to interpret the diagram $L_X$ (or $L_{\rm AGN}$) vs. $M_\star$. The number of objects per logarithmic bins of AGN luminosity and stellar mass is given by
\begin{eqnarray}
\nonumber {{\rm d^2} N\over {\rm d}\log M_{\star}\,{\rm d}\log L_{\rm AGN}} &\simeq & \int{{\rm d}\log\dot M_\star}\,{{\rm d} N\over {\rm d}\log\dot M_\star}\times\\ 
\\
\nonumber &\times & {{\rm d}\delta\over {\rm d}\log L_{\rm AGN}}\,{{\rm d}\delta\over {\rm d}\log M_{\star}}~.
\end{eqnarray}
where the relative times spent by a galaxy with approximately constant SFR in a bin of AGN luminosity and stellar mass are given in Eqs.~(\ref{eq|AGN_duty}) and (\ref{eq|Mstar_duty}). The outcome at the representative redshift $z\approx 2$ is presented in Fig.~\ref{fig|CPmass_z2}, where the color-code indicates the logarithmic number density of objects per unit comoving volume (in Mpc$^{-3}$).
The lilac line with errorbars illustrates the number density-weighted mean relationship $\langle L_X\rangle$ (or $\langle L_{\rm AGN}\rangle$) at given $M_\star$ with its $2\sigma$ variance.

The evolutionary tracks for individual galaxies (dotted lines) feature a spiky behavior. This is because, as can be inferred from Fig.~\ref{fig|lightcurves}, both during the ascending and the declining part of the AGN lightcurve, the AGN luminosity increases or decreases exponentially, while the stellar mass varies little; e.g., before the AGN luminosity peak, the behavior $M_\star\propto \sqrt{\log L_{\rm AGN}}$ applies after Eqs.~(\ref{eq|STARlightcurve}) and (\ref{eq|AGNlightcurve}).
This absence of correlation between $L_{\rm AGN}$ and $M_\star$ is consistent with the observations of X-ray selected AGNs by Mullaney et al. (2012b); as mentioned above, X-ray selection tends to pick up objects both before and after the AGN peak, when the AGN luminosity can be vastly different while the stellar mass changes little.

The mean relationship for detected galaxies (upper lilac dashed line) increases with stellar mass; this is because to achieve a high AGN luminosity, the BH must reside in a galaxy with rather high stellar mass, due to the constraint on the BH to stellar mass ratio at the end of the coevolution, that is pinpointed by a mass-selected sample (cf. Eq.~\ref{eq|BH_Mstar_ratio}). This well agrees with the average relationship for X-ray detected AGNs in the mass-selected galaxy sample by Rodighiero et al. (2015; upper empty circles).

However, the mean relationship when including stacked observations (Mullaney et al. 2015; Rodighiero et al. 2015) lies substantially below, because of the contribution of undetected sources to the average. To compare with these data, it is essential to account for the contribution from star formation to the X-ray emission. In our computation we estimate the X-ray luminosity from star formation by the calibration
\begin{equation}\label{eq|LX_sfr}
L_{X,\rm{SFR}}=7\times 10^{39}\,{\rm erg~s^{-1}}\times {\dot M_\star\over M_\odot~{\rm yr^{-1}}}
\end{equation}
of Vattakunnel et al. (2012).

We stress that the X-ray emission from star formation is usually negligible in individual detected sources with X-ray luminosity $L_X\ga$ a few $10^{42}$ erg s$^{-1}$, corresponding for standard bolometric corrections to $L_{\rm AGN}\ga 10^{44}$ erg s$^{-1}$. Contrariwise, when stacking sources with  detection threshold $L_X\la 10^{42}$ erg s$^{-1}$, the galaxy number density can be largely dominated by starforming objects with negligible nuclear activity. As a consequence, the relation for undetected sources could just mirror the galaxy main sequence, with SFR converted in X-ray luminosity after Eq.~(\ref{eq|LX_sfr}). We find this is indeed the case, with the resulting average relationship for undetected sources (lower lilac dashed line) being a factor about $10^2$ below that for detected AGNs, in agreement with Rodighiero et al. (2015; lower empty circles). We caution that, for such undetected sources, the conversion in bolometric luminosity via the standard X-ray correction for AGNs is somewhat misleading.

The average over detected and stacked sources (lilac solid line) strikes an intermediate course, and is found to be in excellent agreement with the observational determinations by Mullaney et al. (2015) and Rodighiero et al. (2015); however, its straight physical interpretation is hampered because the statistics of detected and undetected objects is contributed by different processes, i.e., either BH accretion or star formation, respectively.

\begin{figure}
\epsscale{1}\plotone{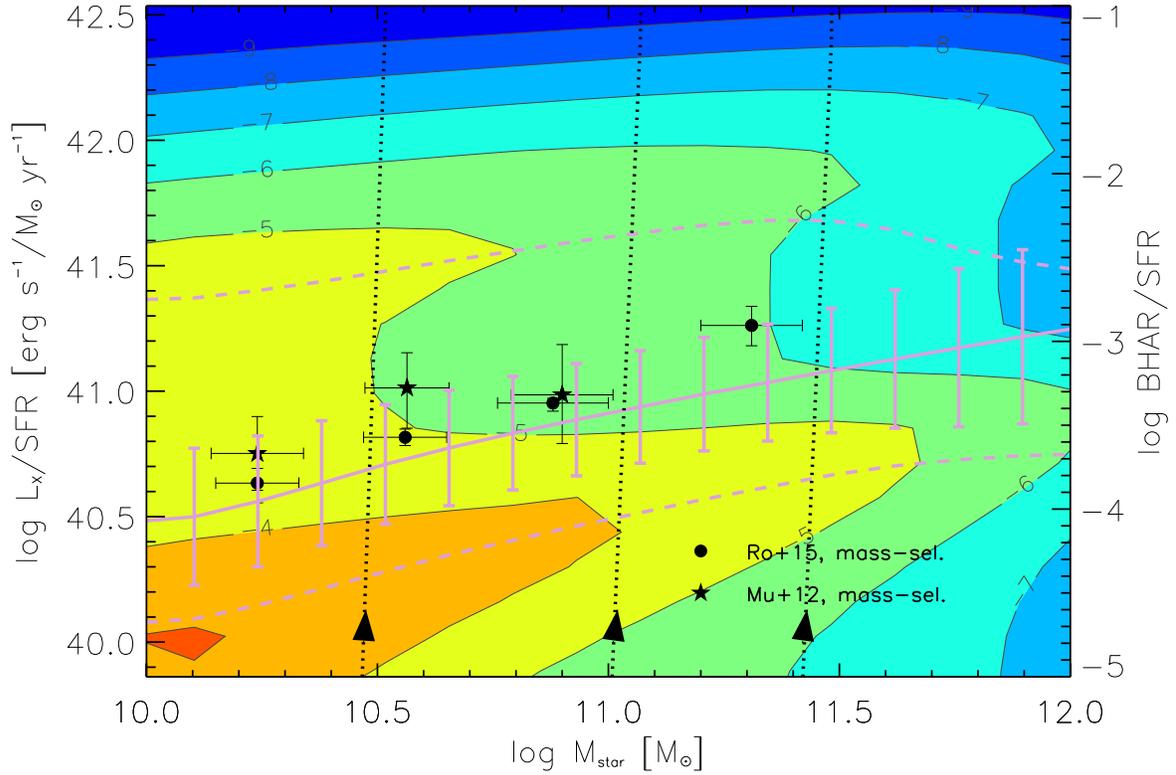}\caption{Relationship between the ratio of the X-ray luminosity (left axis) or the BH accretion rate (right axis, strictly valid only for detected sources) to the SFR vs. the host galaxy stellar mass at $z\approx 2$. The relationship is based on the global (UV+far-IR) SFR functions and the AGN duty cycle from the deterministic BH accretion history of \S~\ref{sec|AGN_history}. Colored contours illustrate the number density of objects (labels are in log units of Mpc$^{-3}$). The lilac line with errorbars illustrates the relationship averaged over detected and stacked sources (dashed lines), with its $2\sigma$ variance, for a detection limit at $L_X\approx 10^{43}$ erg s$^{-1}$. The dotted lines show three evolutionary tracks (forward time direction indicated by arrows) for objects with final stellar mass of about $10^{10.5}$, $10^{11}$, $10^{11.5}\, M_\odot$. Data are from Rodighiero et al. (2015; circles) and Mullaney et al. (2012; stars) for mass-selected samples.}\label{fig|CPmass_ratio_z2}
\end{figure}

In Fig.~\ref{fig|CPmass_ratio_z2} we show the average ratio between the X-ray luminosity (or BH accretion rate) and the SFR vs. the stellar mass. Note that for a fair comparison with observational data by Mullaney et al. (2015) and Rodighiero et al. (2015), the BH accretion rate has been obtained from the X-ray luminosity by converting to bolometric with the standard X-ray correction for AGN and then using $\dot M_{\rm BH}=(1-\epsilon)\, L_{\rm AGN}/\epsilon\, c^2$ with $\epsilon\approx 0.1$. The result is another representation of the previous diagram, so our computation is again based on Eq.~(\ref{eq|AGN_MS}), and the same cautionary comments apply. Specifically, detected sources are truly AGN, while the undetected sources are mainly main sequence galaxies with negligible nuclear activity. The average relationship $L_X/\dot M_\star\propto M_{\star}^{0.5}$ weakly increases with the stellar mass, in agreement with the latest data from Rodighiero et al. (2015).

We stress that such a behavior of the average relationship (mean of detected plus stacked sources) between X-ray luminosity (or BH accretion rate, strictly holding only for detected sources) and the SFR does not mean that the quantities are proportional during the galaxy lifetime. In fact, the evolutionary tracks for individual objects are characterized by a very steep trend; this is because, during the ascending phase, the AGN luminosity increases exponentially, while the SFR is roughly constant and the stellar mass increases mildly with time. It is only in statistical sense that the weakly increasing behavior with $M_\star$ emerges.

Finally, we note that the normalization of the average ratio around a few $10^{-4}$ at $M_\star\approx 10^{11}\, M_\odot$ is driven by the statistics of stacked sources, that is mainly contributed by starforming galaxies with negligible nuclear activity. In fact, the relationship for detected X-ray AGNs lies a factor around $10$ above, with normalization increased to a few $10^{-3}$ at $M_\star\approx 10^{11}\, M_\odot$; indeed this reflects the value of the BH to stellar mass ratio at the end of the coevolution, as pinponted by a mass selection.

\section{Summary and conclusions}\label{sec|summary}

We have provided a novel, unifying physical interpretation on the origin, the average shape, the scatter, and the cosmic evolution for the main sequences of starforming galaxies and AGNs at high redshift $z\ga 1$.
We have achieved this goal in a \emph{model-independent} way by exploiting: (i) the redshift-dependent SFR functions by Mancuso et al. (2016), based on the latest UV and far-IR data from \textsl{HST}, \textsl{Herschel}, and validated on several grounds; (ii) deterministic evolutionary tracks for the history of star formation and black hole accretion, gauged on a wealth of multiwavelength observations. We further validate these ingredients by showing their consistency with the observed galaxy stellar mass functions and AGN bolometric luminosity functions at different redshifts via the continuity equation approach.

Our main findings are as follows:

\begin{itemize}

\item The galaxy main sequence and its scatter originate naturally as a statistical relationship from the SFR functions and a deterministic star formation history. The existence of the main sequence by no means implies that individual galaxies evolve along it. Specifically, we envisage young objects to be preferentially located to the left of the main sequence at given SFR. As the time goes by they will move at nearly constant SFR toward the main sequence locus, spending there most of their lifetime. Afterwards, their SFR will be reduced and the galaxy will move below the main sequence, occupying the locus of 'red and dead' objects, as observed locally. Such a picture provides strong support to an in situ scenario for star formation in galaxies.

\item Off-main sequence galaxies are interpreted as young objects, that have still to accumulate most of their stellar mass. In this respect, the interpretation of their peculiar position in the SFR vs. $M_\star$ diagram is not to be \emph{above} the main sequence at given stellar mass, but rather to be to the \emph{left} of it at given SFR. Mass-selected galaxies tends to preferentially lie on the main sequence, while far-IR selected ones are unbiasedly picked up on it or to the left of it. The age estimates via SED modeling of far-IR selected galaxies support our interpretation, showing a tendency for objects more distant from the main sequence locus to feature smaller and smaller ages. Note that this is in contrast to the interpretation of off-main sequence objects as starbursts triggered by mergers or cosmological inflows.

\item We are also able to explain the redshift evolution of the main sequence toward higher SFRs at given stellar mass, tracing star formation in galaxies back to in situ condensation processes. At higher $z$ these are typically more efficient, yielding more violent SFRs, and so making the main sequence locus to shift upwards. We also expect that going toward higher redshift, the number of off-main sequence objects decreases appreciably.

\item The AGN main sequence is globally understood in terms of star formation and BH accretion histories. As to the SFR vs. $L_{\rm AGN}$ relationship, we expect a flat behavior for AGN luminosities lower than those associated with star formation, reflecting the individual evolutionary tracks of young galaxies. Then on moving toward higher AGN luminosities, the SFR gets rapidly suppressed by feedback processes, and the AGN luminosity itself slowly fades, so that the evolutionary tracks of individual objects move toward the bottom left region of the diagram. Contrariwise, the mean SFR increases with AGN luminosity, being statistically dominated by more massive objects that feature higher and higher SFRs.

\item As to the $L_X$ vs. $M_\star$ relationship, we expect that the evolutionary tracks for individual galaxies feature a spiky behavior, because the AGN luminosity increases or decreases exponentially with the galactic age, while the stellar mass varies little. This absence of correlation is consistent with the observations of X-ray selected AGNs. On the other hand, the mean $L_X$ for detected galaxies increases almost linearly with $M_\star$, being statistically dominated by objects with higher stellar and BH masses. The resulting relation well agrees with the average relationship for X-ray detected AGNs in mass-selected galaxy samples. However, when including stacked observations the mean $L_X$ is substantially lower, because of the contribution of undetected sources. We have highlighted that the statistics of undetected sources is dominated by starforming galaxies with negligible AGN emission, and so it reflects the galaxy main sequence, with SFR converted in X-ray luminosity. As such, a straight physical interpretation of the mean relationship is hampered because the statistics of detected and undetected objects is contributed by different processes, i.e., either BH accretion or star formation, respectively.

\item Finally, we have discussed the increase of the average ratio between BH accretion rate and the SFR as a function of galaxy stellar mass. This does not imply that the BH accretion rate and SFR are proportional during the entire galaxy lifetime, because the average relationship holds only for detected sources in a statistical sense.

\end{itemize}

The star formation and BH accretion histories emerging from the above observational landscape constitute a testbed for galaxy and AGN evolution models; in particular, they provide a guide to gauge recipes in semianalytic models and subgrid physics in numerical simulations.

We have shown that to properly interpret the main sequences for starforming galaxies and AGNs at high redshift $z\ga 2$ future observations should aim at: (i) exploiting different selections for galaxies and AGNs to fully populate the main sequence diagrams, especially for detected sources; (ii) measuring with improved accuracy the ages of stellar populations for extended samples of starforming galaxies and AGN hosts; (iii) determining via spectroscopic observations the gas content of starforming galaxies and AGN hosts located on and above the main sequences. To study off-main sequence objects, an important issue concerns the stellar mass determination of far-IR selected galaxies at high-redshift, that will become feasible in the near future thanks to the \textsl{JWST}.

In the present paper we have provided an unified view on galaxy and supermassive BH evolution. In fact, while there is a large consensus on BH evolution via in situ accretion processes, debate is still open on the role of mergers or cosmological gas inflows in the assembly of stellar mass within galaxies. All in all, our analysis of the main sequence for high-redshift galaxies and AGNs highlights that the present data can be consistently interpreted in terms of an \emph{in situ coevolution} scenario, that strongly demands to be further tested via future observations.

\begin{acknowledgements}
We thank the anonymous referee for a constructive report. We are grateful to F. Fontanot and G. Rodighiero for stimulating discussions. Work partially supported by PRIN INAF 2014 `Probing the AGN/galaxy co-evolution through ultra-deep and ultra-high-resolution radio surveys'. J.G.N. acknowledges financial support from the Spanish MINECO for a ‘Ramon y Cajal’ fellowship (RYC-2013-13256) and the I+D 2015 project AYA2015-65887-P (MINECO/FEDER). CM and AL thank the Astronomy Dept. at USTC in Hefei (China) for the warm hospitality, under the support from the National Science Foundation of China (grant No. 11503024).
\end{acknowledgements}

\clearpage

\clearpage

\clearpage

\clearpage

\clearpage

\clearpage

\clearpage

\clearpage

\clearpage

\clearpage

\clearpage

\clearpage

\clearpage

\clearpage
\begin{turnpage}
\begin{deluxetable}{lcccccccccccccccccccccccccccccccccc}
%\rotate
\tabletypesize{\scriptsize}\tablewidth{0pt}\tablecaption{SFR Function
Parameters}\tablehead{\colhead{Parameter} & &\multicolumn{7}{c}{Far-IR+UV} & & &
\multicolumn{7}{c}{UV (dust-corrected)}\\
\\
\cline{1-1} \cline{3-9} \cline{12-18}\\
\colhead{} & &\colhead{$p_0$} & & \colhead{$p_1$} & & \colhead{$p_2$} & &
\colhead{$p_3$} & & & \colhead{$p_0$} & & \colhead{$p_1$} & & \colhead{$p_2$}
& & \colhead{$p_3$}}\startdata
$\log \mathcal{N}(z)$ & &$-2.53\pm 0.06$& &$-5.81\pm 1.19$& &$10.41\pm 3.57$& &$-5.95\pm 2.59$& & & $-1.90\pm 0.08$& &$-1.50\pm 1.72$& &$4.28\pm 4.46$& &$-5.41\pm 3.02$\\
$\log\dot M_{\star, c}(z)$ & & $1.28\pm 0.05$& &$4.71\pm 0.62$& &$-1.67\pm 1.70$& &$-3.3\pm 1.21$ & & &$-0.04\pm 0.05$& &$1.96\pm 0.99$& &$1.87\pm 2.55$& &$-2.48\pm 1.72$\\
$\alpha(z)$& &$1.29\pm 0.01$& &$2.82\pm 0.23$& &$-6.18\pm 0.67$& &$4.20\pm 0.46$& & &$1.09\pm 0.03$ & & $3.39\pm 0.61$& &$-7.53\pm 1.60$& &$5.41\pm 1.07$\\
\enddata
\tablecomments{Quoted uncertainties are at $1-\sigma$ level. Fits hold in the range of SFR $\dot M_\star\sim 10^{-3}-10^4\, M_\odot$ yr$^{-1}$ and for redshifts $z\sim 0-8$. Here we adopt a SMC extinction law (for a Meurer/Calzetti law, see Table 1 in Mancuso et al. 2016).}
\end{deluxetable}
\end{turnpage}

\end{document}